\documentclass[%
reprint,
superscriptaddress,
 amsmath,amssymb,
 aps,
floatfix,
]{revtex4-2}
\usepackage{mathbbol}
\usepackage{amssymb}   
\usepackage{comment}
\DeclareSymbolFontAlphabet{\amsmathbb}{AMSb}%
\usepackage{graphicx}
\usepackage{dcolumn}
\usepackage{bm}
\usepackage{amsmath,amsfonts,amsthm, mathtools}
\usepackage{xcolor}
\usepackage{braket}
\def\be{\begin{equation}}
\def\ee{\end{equation}}

\newcommand{\op}[2]{|#1\rangle\langle #2|}

\newcommand{\mbb}{\mathbb}
\renewcommand{\ol}{\overline}
\newcommand{\wt}{\widetilde}
\newcommand{\mbf}{\mathbf}
\newcommand{\win}{\text{win}}
\newcommand{\mc}{\mathcal}
\newcommand{\tr}{\text{Tr}}
\newcommand{\msf}{\mathsf}

\theoremstyle{definition}

\newtheorem{proposition}{Proposition}
\newtheorem{theorem}{Theorem}

\newtheorem*{remark}{Remark}

\usepackage{color}
\definecolor{cool_green}{rgb}{0.0, 0.5, 0.0}
\newcommand{\eric}[1]{{\color{cool_green} #1}}

\begin{document}

\preprint{APS/123-QED}

\title{Quantum Telescopy Clock Games}


\author{Robert Czupryniak}
\email{rczupryn@ur.rochester.edu}
\affiliation{Department of Physics and Astronomy, University of Rochester, Rochester, NY 14627}
\affiliation{Center for Coherence and Quantum Optics, University of Rochester, Rochester, NY 14627}
\affiliation{Institute for Quantum Studies, Chapman University, Orange, CA 92866}
\author{Eric Chitambar}
\affiliation{Department of Electrical and Computer Engineering, Coordinated Science Laboratory, University of Illinois at Urbana-Champaign, Urbana, Illinois 61801, USA}
\author{John Steinmetz}
\affiliation{Department of Physics and Astronomy, University of Rochester, Rochester, NY 14627}
\affiliation{Center for Coherence and Quantum Optics, University of Rochester, Rochester, NY 14627}
\affiliation{Institute for Quantum Studies, Chapman University, Orange, CA 92866}
\author{Andrew N. Jordan}
\affiliation{Department of Physics and Astronomy, University of Rochester, Rochester, NY 14627}
\affiliation{Center for Coherence and Quantum Optics, University of Rochester, Rochester, NY 14627}
\affiliation{Institute for Quantum Studies, Chapman University, Orange, CA 92866}

\begin{abstract}
We consider the clock game-a task formulated in the framework of quantum information theory-that can be used to improve the existing schemes of quantum-enhanced telescopy. The problem of learning when a stellar photon reaches a telescope is translated into an abstract game, which we call the clock game.  A winning strategy is provided that involves performing a quantum non-demolition measurement that verifies which stellar spatio-temporal modes are occupied by a photon without disturbing the phase information.  We prove tight lower bounds on the entanglement cost needed to win the clock game, with the amount of necessary entangled bits equaling the number of time-bins being distinguished.  This lower bound on the entanglement cost applies to any telescopy protocol that aims to non-destructively extract the time-bin information of an incident photon through local measurements, and our result implies that the protocol of Khabiboulline \textit{et al.} [\text{Phys. Rev. Lett.} \textbf{123}, 70504 (2019)] is optimal in terms of entanglement consumption.  The full task of the phase extraction is also considered, and we show that the quantum Fisher information of the stellar phase can be achieved by local measurements and shared entanglement without the necessity of nonlinear optical operations. The optimal phase measurement is achieved asymptotically with increasing number of ancilla qubits, whereas a single qubit pair is required if nonlinear operations are allowed.
\end{abstract}

\maketitle

\tableofcontents

\section{Introduction}

Quantum games offer a quantitative framework to isolate and study different features of quantum mechanics.  The most well-known type of game studied in the literature are nonlocal games \cite{Cleve-2004a, Buscemi-2012b, Johnston-2016a}, which capture the properties of entanglement that cannot be described by local hidden variable models \cite{Brunner-2014a}.  Other types of games have been proposed to characterize features of statistical comparisons \cite{Buscemi-2012a}, wave-particle duality and quantum coherence \cite{Napoli-2016a, Coles-2016a, Biswas-2017a, Bagan-2018a, Del-Santo-2020a, Zhang-2020a}, quantum steering \cite{Wiseman-2007a}, measurement incompatibility \cite{Skrzypczyk-2019a, Carmeli-2019a, Buscemi-2020a}, and general resource theories \cite{Takagi-2019a, Uola-2019a}.  In this paper we invoke the notion of quantum games to study the problem of nonlocal phase estimation. 

\begin{figure}
    \centering
    \includegraphics[width=6cm]{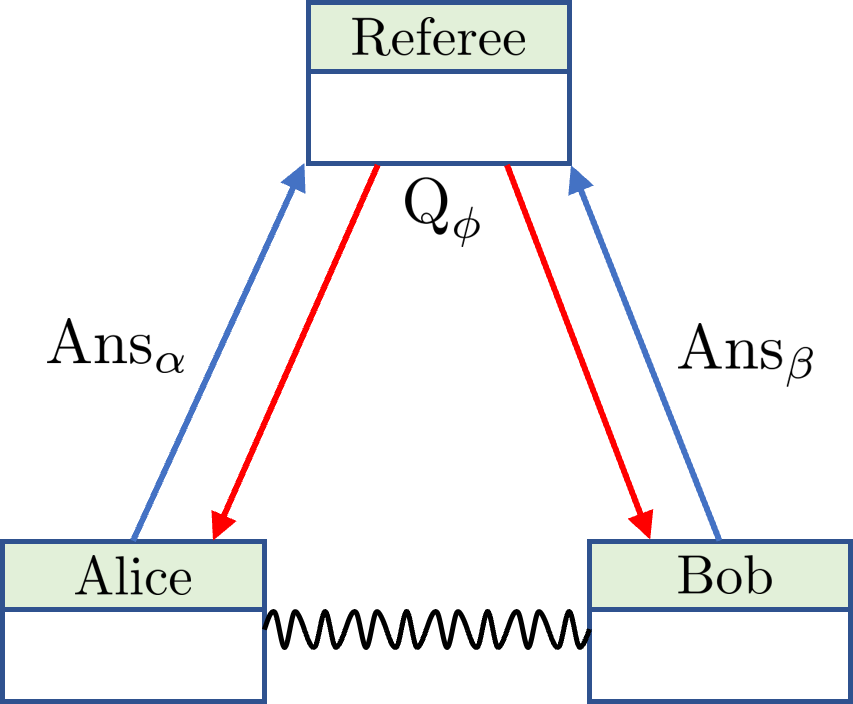}
    \caption{A general bipartite quantum game consists of questions and answers between a referee and two non-communicating parties.  However, the parties can use shared entanglement (wavy line) to coordinate their responses.}
    \label{fig:game}
\end{figure}

A general bipartite quantum game consists of two players (Alice and Bob) and a referee (see Fig. \ref{fig:game}).  The referee asks Alice and Bob some question $\text{Q}_\phi$, which in general consists of both quantum and classical parts.  Alice and Bob then return answers, $\text{Ans}_\alpha$ and $\text{Ans}_\beta$ respectively, that again may have both quantum and classical parts.  While Alice and Bob are not able to communicate classically when formulating their answers, they do have access to some shared entanglement which they can use to coordinate their answers.  Each game has some winning condition in terms of what Alice and Bob should return for a given question, and their goal is to devise a strategy that maximizes the probability of winning.

In the game we consider, $\text{Q}_\phi$ consists of a phase-encoded entangled state $(\ket{1_A 0_B}+e^{i\phi}\ket{0_A 1_B})/\sqrt{2}$ that the referee distributes to Alice and Bob in time-bin $n\in\{1,2,\cdots, N\}$, while the vacuum is received from the referee within all other time-bins. They win the game if they reply with classical data that correctly identifies time-bin $n$ along with a bipartite state that possesses the same relative phase $\phi$. Hence the overall objective is to nonlocally extract some classical information about the phase-encoded state (its time-bin) without disturbing its phase information.

One motivation for considering this game comes from the task of quantum long-baseline telescopy \cite{gottesman2012longer}. The quantum-enhanced version of very-long-baseline interferometry (VLBI) refers to the method of imaging stellar objects by collecting emitted photons at spatially-separated telescopes and studying their interference profile \cite{Monnier-2003a}.  Here the qubit states $\ket{0}$ and $\ket{1}$ within the phase-encoded state correspond to the vacuum and a single photon in a given spatio-temporal mode. Directly transferring the remotely captured photons to a single interferometer can be challenging due to noise and loss, but quantum mechanics offers an alternative solution. As first proposed by Gottesman \textit{et al.} \cite{gottesman2012longer}, the physical transfer of stellar photons from each telescope to a central station can be replaced by a network of quantum repeaters that distributes an entangled state to telescope locations. The original scheme by Gottesman \textit{et al.} requires a very large entanglement generation rate between the two telescopes, but a modified protocol that use quantum memories has recently been proposed that is less demanding in terms of its entanglement cost \cite{khabiboulline2019optical, Khabiboulline-2019b}.  A key property of this new protocol is that it effectively decouples the time-bin information of an incoming stellar photon from its phase information.  Our game can thus be seen as a full abstraction of this idea in which the star is replaced by a photon-distributing referee, and the primary goal is for Alice and Bob to collect the time-bin data without disrupting the phase.  In principle an optimal strategy for Alice and Bob in this game could be used as a subroutine in a large phase estimation protocol.

One advantage of adopting this abstract approach is that it allows us to evaluate the quality of different phase estimation protocols using game-theoretic measures beyond just the standard quantifier of Fisher information. That way we can not only quantify the amount of information one gains in each quantum measurement, but also consider the amount of resources (e.g., entanglement cost) needed for the measurement scheme. In addition, by formulating the various components of a phase estimation protocol in terms of a nonlocal game, we can analyze trade-offs between winning success probabilities and the entangled resources that Alice and Bob use in the game. A primary objective of this paper is to construct and analyze new phase estimation protocols that use different forms of shared entanglement between Alice and Bob.  One of our main results (Theorem \ref{Thm:LOSE-entanglement}) places a tight lower bound on the entanglement needed to non-destructively extract the time-bin information through local measurements.  Hence, any distributed telescopy protocol that involves decoupling the time-bin and phase information, such as those in \cite{khabiboulline2019optical, Khabiboulline-2019b, CNOT_scheme}, will require this much entanglement.  Since the telescopy protocol first presented in \cite{khabiboulline2019optical} saturates this lower bound, we have proved its optimality in terms of entanglement cost.

The structure of this paper is as follows: In Sec. \ref{sec:clock_game} we introduce the clock game and propose the winning strategy. We examine the resources needed to win the game, which includes the ancilla quantum state that contains a certain degree of entanglement. We study what conditions the ancilla state must satisfy to win the game with certainty and quantify how the errors introduced to the ancilla state reduce the winning probability. Sec. \ref{Sect:linear-phase-extraction} introduces the phase extraction protocol that performs an optimal measurement of the stellar phase without the necessity of nonlinear optical elements. We include an analysis of the resources required to perform that protocol. In Sec. \ref{sec:conclusions} we conclude. 

\section{Clock game} \label{sec:clock_game}

\begin{figure}
    \centering
    \includegraphics[width=8.6cm]{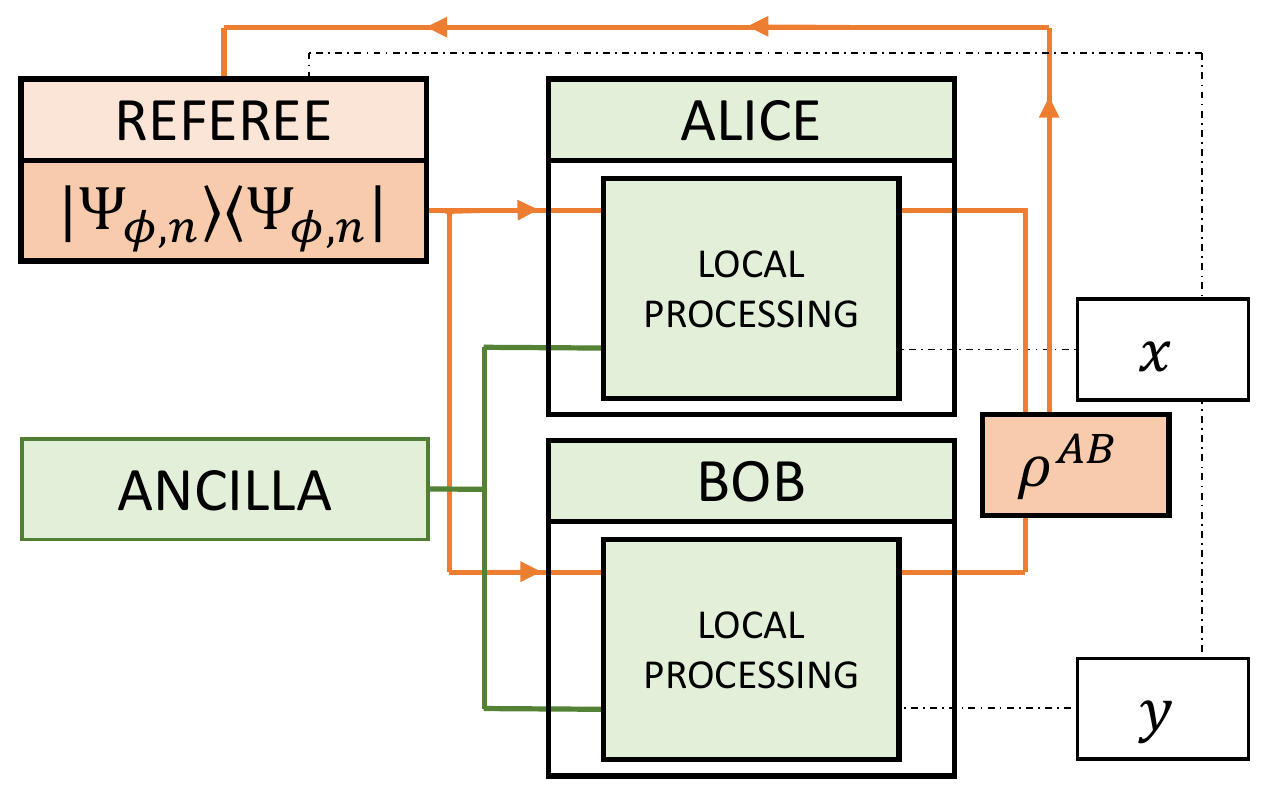}
    \caption{Schematic representation of the clock game. The referee delivers to Alice and Bob the phase encoded state $\ket{\Psi_{\phi,n}}$ encoded in $2N$ qubits; each party receive half of them. They also receive an ancilla quantum state which they are free to specify. Both parties are allowed to manipulate the locally available quantum states to extract two pieces of classical information: integers $x$ and $y$. As a result, the qubits received from the referee are modified to the state $\rho^{AB}$. Alice and Bob send $(x,y,\rho^{AB})$ back to the referee. }
    \label{fig:clock_prot_summary}
\end{figure}

The clock game is summarized in Fig. \ref{fig:clock_prot_summary}. Alice, Bob and the referee are in different physical locations. The rules of the game are as follows:
\begin{enumerate}
    \item The referee sends a phase-encoded state
    \begin{align} \label{eq:ref_bell_pair}
     \ket{\Psi_{\phi,n}}=\frac{\ket{1,n}_{A}\ket{0,n}_{B}+e^{i\phi}\ket{0,n}_{A}\ket{1,n}_{B}}{\sqrt{2}}
    \end{align}
    to Alice and Bob, where $\ket{j,n}$ denotes $j$ excitations in the $n^{th}$ time-bin and no excitations in the other time-bins. Note that $\ket{\Psi_{\phi,n}}$ can be considered as an element of $\mbb{C}^3\otimes\mbb{C}^N$, a space spanned by vectors $\{\ket{1,n}_A\ket{0,n}_B,\ket{0,n}_A\ket{1,n}_B,\ket{0,n}_A\ket{0,n}_B\}_{n=1}^N$.  The indices $A$ and $B$ indicate the qubits sent to Alice and Bob, respectively. Only the referee knows both $n$ and $\phi$. The set of possible time-bins $\{1, 2, ..., N\}$ is known to all parties. \\
    
    Alternatively, the referee can trick Alice and Bob by not sending the state (\ref{eq:ref_bell_pair}) at all and send the vacuum within all the time bins. In that case we will use the index $n=0$. Equation (\ref{eq:ref_bell_pair}) is valid for indices $n>0$, and $\ket{\Psi_{\phi,0}}$ denotes the vacuum within all possible time-bins.
    \item Alice and Bob process the data sent from the referee along with some ancilla systems. The ancilla systems are allowed to be entangled states shared by both parties. Alice and Bob are free to specify which ancilla states they receive, including qudit states. Any local processing of $\ket{\Psi_{\phi,n}}$ and the ancilla states is then allowed. 
    \item Alice and Bob reply to the referee with data $(x,y,\rho_{x,y}^{AB})$.  The values $x,y\in\{0,\cdots,N\}$ are classical data sent from Alice and Bob, respectively, and $\rho_{x,y}^{AB}$ is the quantum state received by the referee after Alice and Bob process $\ket{\Psi_{\phi,n}}$. 
    \item The referee measures $\rho_{x,y}^{AB}$ using the projective measurement $\{\ket{\Psi_{\phi,n}}\bra{\Psi_{\phi,n}},\mathbb{1} -\ket{\Psi_{\phi,n}}\bra{\Psi_{\phi,n}}\}$. Alice and Bob win if $n=(x+y)\mod N+1$ and if the referee gets outcome $\ket{\Psi_{\phi,n}}$ in the measurement.
    
    $*$ 
    If the referee has not supplied the phase-encoded state, then Alice and Bob should send classical responses such that $0=x+y\mod N+1$.  In this case, the referee measures $\rho_{x,y}^{AB}$ with projective measurement $\{\ket{\Psi_{\phi,0}}\bra{\Psi_{\phi,0}},\mathbb{1} -\ket{\Psi_{\phi,0}}\bra{\Psi_{\phi,0}}\}$. 
\end{enumerate}

As in all nonlocal games, Alice and Bob are not allowed to communicate during this protocol, although they can make use of shared randomness and entangled ancilla to coordinate their actions.  Formally then, any strategy that Alice and Bob employ can be characterized by a local operations and shared entanglement (LOSE) instrument $\{\mc{L}_{x,y}\}_{x,y=0}^N$ \cite{Buscemi-2012b}.  This is a collection of completely positive (CP) maps such that each $\mc{L}_{x,y}$ can be expressed as
\begin{equation}
    \mc{L}_{x,y}(\Psi^{AB})=\sum_\lambda p(\lambda) \mc{A}^{AA'}_{x|\lambda}\otimes\mc{B}^{BB'}_{y|\lambda}(\Psi^{AB}\otimes \varphi^{A'B'}),
\end{equation}
where $\varphi^{A'B'}$ is some fixed entangled ancilla, and both $\sum_{x=0}^N\mc{A}^{AA'}_{x|\lambda}$ and $\sum_{y=0}^N\mc{B}^{BB'}_{y|\lambda}$ are  trace-preserving for every $\lambda$.  For an input state $\Psi\in\mc{D}(\mbb{C}^3\otimes\mbb{C}^N)$, Alice and Bob obtain the classical output $(x,y)$ with probability $p(x,y):=\tr[\mc{L}_{x,y}(\Psi)]$, and their post-measurement state is $\rho_{x,y}:=\mc{L}_{x,y}(\Psi)/{p(x,y)}$.  Both the classical and quantum outputs of the instrument are forwarded to the referee.  If the referee encodes phase $\phi$ in time-bin $n$, the probability that Alice and Bob win using an instrument $\{\mc{L}_n\}_{x,y=0}^N$ is given by
\begin{align}
    P_{\win}(\phi,n)&=
        p(x,y)\bra{\Psi_{\phi,n}}\rho_{x,y}\ket{\Psi_{\phi,n}}\delta_{n,\text{$x+y$ mod $D$}}\notag\\
    &=\bra{\Psi_{\phi,n}}\mc{L}_{x,y}(\Psi_{\delta,n})\ket{\Psi_{\phi,n}}\delta_{n,\text{$x+y$ mod $D$}}.
\end{align}
It is assumed that the referee chooses $n$ and $\phi$ uniformly from the sets $\{0,1,\cdots,N\}$ and $[0,2\pi)$, respectively.  For a given strategy, the winning probability for Alice and Bob is then 

\begin{align}
    P_{\win}=\frac{1}{N+1}\sum_{n=0}^{N}\int_{0}^{2\pi} d\phi P_{\win}(\phi,n).
\end{align}

In the following sections, we provide a winning strategy for the clock game that uses a qudit entangled ancilla state.  Note that any bipartite qudit state is locally equivalent to multiple qubit states, and so our winning strategy can also be seen as a multi-qubit protocol.  In Secs. \ref{sec:multi1}-\ref{sec:multiparty_singleQudit} we generalize the protocol to the multi-party scenario.

\subsection{Elements of qudit computation formalism}

Before proceeding to the analysis of the game, we need some elements of qudit computation formalism. We will use 
\begin{equation}
    \{ \ket{0}, \ket{1}, ... , \ket{D-1} \}
\end{equation}
as the computational basis describing the states of a $D$-level system. The following vectors
\begin{equation} \label{eq:Fourier_basis}
    \ket{\tilde{j}}: = \frac{1}{\sqrt{D}}\sum_{k=0}^{D-1} \exp\left(\frac{2\pi i \tilde{j} k}{D}  \right)\ket{k}.
\end{equation}
form the Fourier basis. In (\ref{eq:Fourier_basis}) the allowed values of $\tilde{j}$ are $0,1,...,D-1$, and the inverse relation is 
\begin{equation}
    \ket{j} = \frac{1}{\sqrt{D}}\sum_{\tilde{k}=0}^{D-1} \exp\left(-\frac{2\pi i j \tilde{k}}{D}  \right) \ket{\tilde{k}} .
\end{equation}
We introduce the qudit Z gate \cite{wang2020qudits}
\begin{equation}
    \hat{Z} \ket{ j } = \exp\left(\frac{2\pi i j}{D}  \right)\ket{ j }.
\end{equation}

The qubit symmetric Bell state $(\ket{00}+\ket{11})/\sqrt{2}$ can be generalized to the qudit case
\begin{equation} \label{eq:qudit_bell_state}
    \ket{\Phi_{D,0}^{(2)}} = \frac{1}{\sqrt{D}}\sum_{j=0}^{D-1} \ket{j}_1\otimes\ket{j}_2.
\end{equation}
Analogously, the generalization of the GHZ state is
\begin{equation}
    \ket{\Phi_{D,0}^{(K)}} 
     = \frac{1}{\sqrt{D}}\sum_{j=0}^{D-1} \ket{j}_1\otimes\ket{j}_2\otimes...\otimes\ket{j}_K.
\end{equation}

Finally, we introduce the controlled-$Z^n$ gate denoted by $CZ^n$, for which a qubit serves as a control and a qudit serves as a target. The gate acts according to the following rules

\begin{equation} \label{eq:CZn_gate}
\begin{aligned}
    U [ CZ^n ] \ket{0}_c \otimes \ket{j}_t 
    & = \ket{0}_c \otimes \ket{j}_t  \\
    U [ CZ^n ] \ket{1}_c \otimes \ket{j}_t 
    & = \ket{1}_c \otimes Z^n \ket{j}_t, 
\end{aligned}
\end{equation}
where by the indices $c$ and $t$ we denote the control qubit and target qudit respectively. $U [ CZ^n ]$ is the unitary operator that applies the $CZ^n$ gate.

\subsection{Winning strategy} \label{sec:pairwise_1pair}

\begin{figure}
    \centering
    \includegraphics[width=8.6cm]{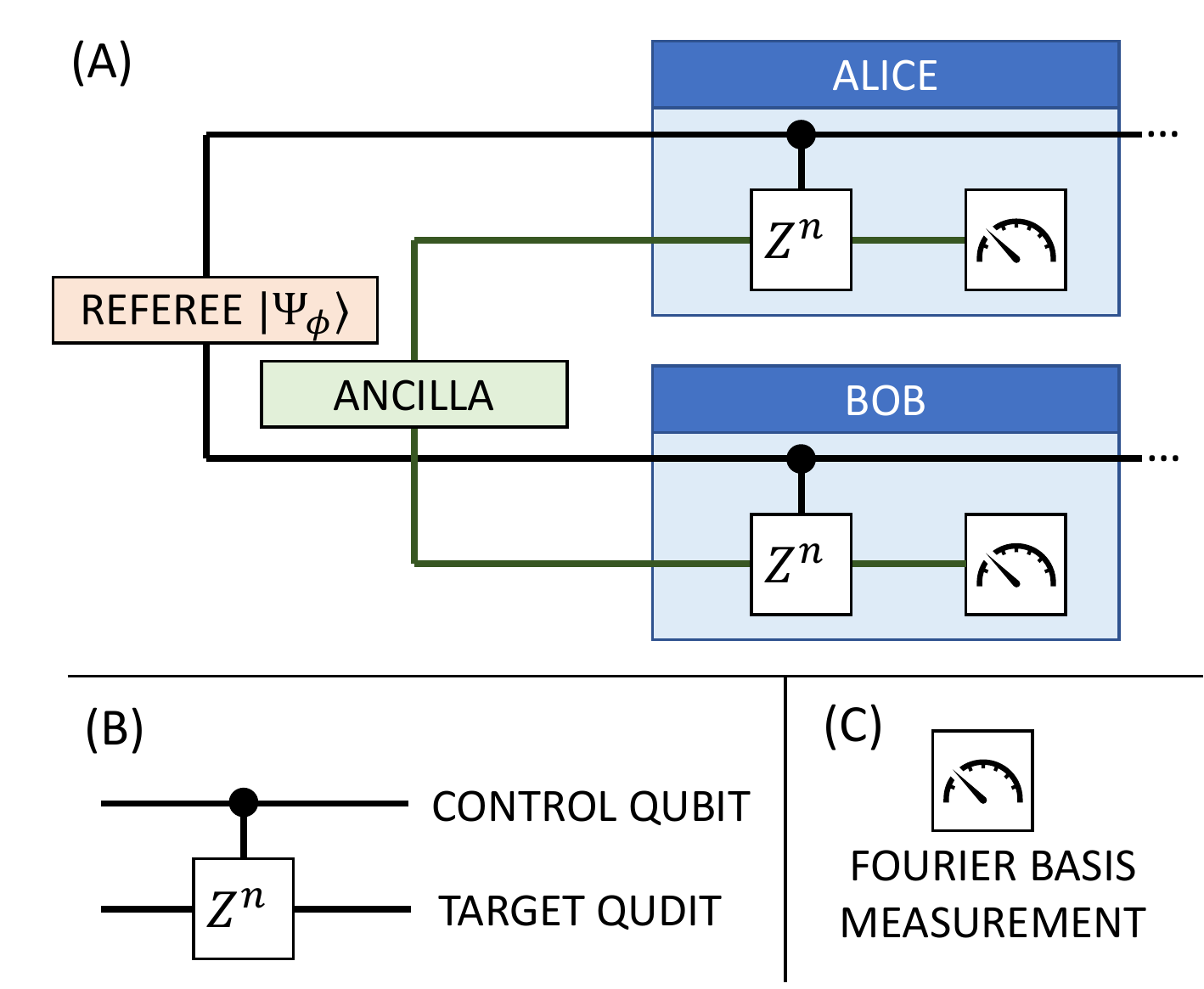}
    \caption{(A) Circuit representation of the operations performed by Alice and Bob in the clock game. This procedure is later followed by sending the answer data $(x,y, \rho_{x,y}^{AB})$ back to the referee. (B) $CZ^n$ gate representation. (C) Symbol for Fourier basis measurement.}
    \label{fig:pairwise_circuit}
\end{figure}

We allow Alice and Bob to share the qudit ancilla state (\ref{eq:qudit_bell_state}), where each party receives one qudit. As we prove below, this allows Alice and Bob to examine at most $N=D-1$ time-bins, where $D$ is the number of levels in each qudit.

The time-bin decoding procedure is described in Fig. \ref{fig:pairwise_circuit}. Suppose that the referee has supplied the phase-encoded state $\ket{\Psi_{\phi,n}}$. 
This provides the control systems for the $CZ^n$ gates located within the local laboratories. The ancilla qudits serve as targets. 
To understand how the referee's state would affect the ancilla qudits, we observe the following property  describing the $Z^n$ gate and the $\ket{\Phi_{D,0}^{(2)}}$ state
\begin{equation} \label{eq:Zn_effect}
    \left[ Z^n \otimes \mathbb{1} \right]\ket{\Phi_{D,0}^{(2)}}
    = \left[ \mathbb{1} \otimes Z^n \right]\ket{\Phi_{D,0}^{(2)}}
    \equiv \ket{\Phi_{D,n}^{(2)}}
\end{equation}
so it does not matter if the $Z^n$ gate acts on the first (Alice's) or second (Bob's) qudit in $\ket{\Phi_{D,0}^{(2)}}$; the resulting state is the same. The state $\ket{\Phi_{D,n}^{(2)}}$ can be expressed in the Fourier basis as
\begin{equation} \label{eq:phi_n_2}
    \ket{\Phi_{D,n}^{(2)}} = \frac{1}{\sqrt{D}} \sum_{\tilde{x}=0}^{D-1} \sum_{\tilde{y}=0}^{D-1} \delta_{\tilde{x}+\tilde{y},n}^{(D)} \ket{\tilde{x}} \otimes \ket{\tilde{y}},
\end{equation}
where we use the following variation of the Kronecker delta function
\begin{equation} \label{eq:mod_delta}
    \delta_{\tilde{x}+\tilde{y},n}^{(D)} = 
    \begin{cases}
        1 \quad  & \text{if} \quad (\tilde{x}+\tilde{y})\text{ mod }D = n \\
        0 & \text{otherwise}.
    \end{cases}
\end{equation}

To win the game, both Alice and Bob perform the $U[CZ^n]$ on the locally available quantum states. The ancilla qubits are modified only within the time-bin occupied by the phase-encoded referee qubit pair. Within that time-bin, the ancilla is modified according to

\begin{equation}
\begin{aligned}
    & U_A[CZ^n] \otimes U_B[CZ^n] \left( \ket{\Psi_{\phi,n}'} \otimes \ket{\Phi_{D,0}^{(2)}} \right) \\
    & = \frac{1}{\sqrt{2}} U_A[CZ^n] \ U_B[CZ^n] \ket{1}_{A} \ket{0}_{B} \ket{\Phi_{D,0}^{(2)}} \\
    & + \frac{e^{i\phi}}{\sqrt{2}} U_A[CZ^n] \ U_B[CZ^n] \ket{0}_{A} \ket{1}_{B} \ket{\Phi_{D,0}^{(2)}} \\
    & = \frac{1}{\sqrt{2}} \ket{1}_{A} \ket{0}_{B} \ \left[ Z^n \otimes \mathbb{1} \right] \ket{\Phi_{D,0}^{(2)}} \\
    & + \frac{e^{i\phi}}{\sqrt{2}} \ket{0}_{A} \ket{1}_{B} \ \left[ \mathbb{1} \otimes Z^n \right] \ket{\Phi_{D,0}^{(2)}} \\
    & = \frac{1}{\sqrt{2}} \ket{1}_{A } \ket{0}_{B } \ket{\Phi_{D,n}^{(2)}} + \frac{e^{i\phi}}{\sqrt{2}} \ket{0}_{A } \ket{1}_{B } \ket{\Phi_{D,n}^{(2)}} \\
    & = \ket{\Psi_{\phi,n}'} \otimes \ket{\Phi_{D,n}^{(2)}}
    \label{Eq:CZCZ}
\end{aligned}
\end{equation}
where by $\ket{\Psi_{\phi,n}'}$ we indicated only a pair of referee qubits that has the phase encoded in it
\begin{equation}
    \ket{\Psi_{\phi,n}'} = \frac{\ket{1}_{A}\ket{0}_{B}+e^{i\phi}\ket{0}_{A}\ket{1}_{B}}{\sqrt{2}}.
\end{equation} 
The indices $A$ and $B$ denote the $CZ^n$ gates performed by Alice and Bob, respectively. Starting from the second line in Eq. \eqref{Eq:CZCZ}, we have omitted some tensor product signs. Note that the resulting state is a separable state of the referee qubits and ancilla qudits, where the ancilla quantum state $\ket{\Phi_{D,n}^{(2)}}$ has the time-bin $n$ encoded in it. It has the important property
\begin{equation}
    \braket{\Phi_{D,n}^{(2)} | \Phi_{D,m}^{(2)}} = \delta_{n,m}
\end{equation}
where the object on the right is the standard Kronecker delta function. It ensures that sending the entangled pair by the referee in different time-bins will result in well-distinguishable ancilla states. That also provides the reason for the choice $N\leq D-1$; the procedure given above assigns one of the $D$ states of the ancilla to each time bin. $D-1$ of them correspond to different time-bins within which the referee can send the phase-encoded state, and the remaining state is used to detect the case of the referee sending the vacuum state. 

The next step is the decoding of the time-bin $n$. After both parties perform all of the $CZ^n$ gates, they perform measurements of the locally available ancilla qudits in the Fourier basis (\ref{eq:Fourier_basis}) and obtain the results $x$ and $y$, which they sent to the referee. These results obey

\begin{equation}
    n' = (x + y) \text{ mod }D.
\end{equation}
According to equations (\ref{eq:phi_n_2}) and (\ref{eq:mod_delta}), it should return the time-bin within which the referee has provided the entangled pair. If both parties have not received a state from the referee at all, then one obtains $n'=0$. Finally, they send the referee's qubits back to her, since the procedure has left the referee's state $\ket{\Psi_{\phi,n}}$ unmodified. Therefore, the projective measurement performed by the referee must return the right result. 
This completes the task.

\subsection{Errors in the ancilla state} \label{sec:errors}

Under ideal conditions, the previous protocol will enable Alice and Bob to learn the time-bin $n$ without disturbing the phase.  However, in realistic conditions their success probability will be bounded away from one.  In particular, interactions with the environment can cause amplitude damping and dephasing errors.  To analyze how this affects the winning probability, we will assume there are no problems with the referee's state preparation and focus exclusively on these types of errors in the ancilla.


\subsubsection{Amplitude damping}
A qudit can experience relaxation between any pair of levels, but the most significant source of these errors is between each adjacent pair of levels $(m,m+1)$. This decay process is governed by the master equation
\begin{equation}
    \frac{d\rho}{dt}=\sum_i\sum_mL_{i,m} \rho L_{i,m}^\dagger-\frac{1}{2}(L_{i,m}^\dagger L_{i,m} \rho + \rho L_{i,m}^\dagger L_{i,m}),
\end{equation}
where $L_{i,m}=\sqrt{\Gamma^{(1)}_{i,m}}\ket{m}\bra{m+1}$ are Lindblad operators acting on qudit $i=1,2$ and $\Gamma^{(1)}_{i,m}$ is the decay rate between levels $(m,m+1)$ of qudit $i$. We use this master equation to find the ancilla state after a time interval $\Delta t$,

\be 
\begin{split}
\rho_{D,0}^{(2)}&= \frac{1}{D}\sum_{j,k}\ket{j,j}\bra{k,k} \\
&+\frac{\Delta t \Gamma^{(1)}_{1,m}}{D}\sum_m\ket{m,m+1}\bra{m,m+1} \\
&+\frac{\Delta t \Gamma^{(1)}_{2,m}}{D}\sum_m\ket{m+1,m}\bra{m+1,m} \\
&-\frac{\Delta t(\Gamma^{(1)}_{1,m}+\Gamma^{(1)}_{2,m})}{2D}\sum_{m,n}\big(\ket{m+1,m+1}\bra{n,n}+\text{h.c.}\big).
\end{split}
\ee

If we apply $Z^n$ to either qudit, we obtain the same result,
\begin{equation}
\begin{aligned}
    \rho_{D,0}^{(2)} \rightarrow \rho_{D,n}^{(2)}
    & = \left[ Z^n \otimes \mathbb{1} \right] \rho_{D,0}^{(2)}\left[ (Z^\dagger)^n \otimes \mathbb{1} \right] \\
    & = \left[ \mathbb{1} \otimes Z^n \right] \rho_{D,0}^{(2)}\left[\mathbb{1} \otimes (Z^\dagger)^n \right].
\end{aligned}
\end{equation}
We then rewrite this in the Fourier basis (for simplicity, we only write the diagonal terms),
\begin{equation}
\begin{aligned}
   \rho_{D,n}^{(2)}(\text{diag.}) 
   & = \frac{1}{D}\sum_{\tilde{p},\tilde{q}}\ket{\tilde{p}\tilde{q}}\bra{\tilde{p}\tilde{q}} \\
   & \times \left[\delta^{(D)}_{\tilde{p}+\tilde{q},n}\left(1-\frac{\Delta t\Gamma^{(1)}}{D}\right)+\frac{\Delta t\Gamma^{(1)}}{D^2}\right],
\end{aligned}
\end{equation}
where $\Gamma^{(1)}=\sum_{i,m}\Gamma^{(1)}_{i,m}$ is the total decay rate of the system. Compared with the ideal result \eqref{eq:phi_n_2}, there are extra terms depending on the decay rate. The win probability is the sum of the diagonal density matrix elements corresponding to $\bar{p}+\bar{q} \text{ mod } N+1=n$, which is
\begin{equation}
    P_{\text{win}}=1-\frac{\Delta t\Gamma^{(1)}(D-1)}{D^2}.
\end{equation}
The rest of the time, the game is lost because $\bar{p}+\bar{q} \text{ mod }N+1\neq n$, where there is equal probability of returning any incorrect time-bin. If the referee tries to trick Alice and Bob by sending the vacuum state in every time bin, then the win probability is reduced by the same amount. Since we linearized the master equation to obtain this result, it is only accurate for small time intervals, i.e., $\Delta t \Gamma^{(1)}\ll 1$. This is a safe assumption since the decay rate due to spontaneous emission should be much slower than the time needed to implement the protocol.

\subsubsection{Dephasing}
We can follow the same steps to find the win probability if the ancilla state has undergone dephasing. We use the same master equation, but with $L_{i,m}=\sqrt{\Gamma^{(2)}_{i,m}/2}\ket{m}\bra{m}$, where $\Gamma^{(2)}_{i,m}$ is the dephasing rate associated with the $m$th level of the $i$th qudit. Note that there are $D$ Lindblad operators for qudit dephasing, as opposed to $D-1$ for amplitude damping. After undergoing dephasing for a time $\Delta t$, the ancilla is
\be 
\begin{split}
\rho_{D,0}^{(2)}&=\frac{1}{D}\sum_{j,k}\ket{j,j}\bra{k,k} \\
&+\frac{\Delta t}{D}\sum_m\frac{\Gamma_{1,m}^{(2)}+\Gamma_{2,m}^{(2)}}{2}\bigg[\ket{m,m}\bra{m,m} \\
&-\frac{1}{2}\sum_n\big(\ket{m,m}\bra{n,n}+\ket{n,n}\bra{m,m}\big)\bigg].
\end{split}
\ee

Once again, we get the same result $\rho^{(2)}_{D,n}$ by applying $Z^n$ to either qudit, and write the diagonal terms in the Fourier basis,
\begin{equation}
\begin{aligned}
    \rho^{(2)}_{D,n}(\text{diag.})
    & = \frac{1}{D}\sum_{\tilde{p},\tilde{q}}\ket{\tilde{p}\tilde{q}}\bra{\tilde{p}\tilde{q}} \\
    & \times \left[\delta^{(D)}_{\tilde{p}+\tilde{q},n}\left(1-\frac{\Delta t\Gamma^{(2)}}{2D}\right)+\frac{\Delta t\Gamma^{(2)}}{2D^2}\right],
\end{aligned}
\end{equation}
where $\Gamma^{(2)}=\sum_{i,m}\Gamma^{(2)}_{i,m}$ is the total dephasing rate. The win probability in this case is
\begin{equation}
    P_{\text{win}}=1-\frac{\Delta t\Gamma^{(2)}(D-1)}{2D^2},
\end{equation}
and once again, if the protocol fails then it has equal probability of returning any of the incorrect time bins. The same is true if the referee sends the vacuum state in every time bin. Combining the effects of amplitude damping and dephasing gives the win probability
\begin{equation}
    P_{\text{win}}=1-\frac{\Delta t(D-1)}{D^2}\left(\Gamma^{(1)}+\frac{\Gamma^{(2)}}{2}\right).
\end{equation}

\subsection{Entanglement Cost under General LOSE} \label{sec:LOSE}

The winning strategy for the clock game presented in Sec. \ref{sec:pairwise_1pair} involved Alice and Bob simply performing local unitaries.  But in principle they could perform more general operations if they use local ancilla systems in addition to the shared entangled ancilla system.  In this section we examine the amount of entanglement needed to win the clock game using the most general local strategy.  Since they are allowed to have shared entanglement and randomness as a resource, we must consider the problem within the framework of LOSE transformations.  Ultimately we will find that the local unitary protocol of Sec. \ref{sec:pairwise_1pair} is optimal in terms of entanglement consumption.

We are interested in understanding LOSE transformations of the form
\begin{align}
\label{Eq:LOSE-transform}
    \Psi_{\phi,n}^{AB}\otimes\varphi^{A'B'}\mapsto\Psi_{\phi,n}^{AB}\otimes\sum_{\substack{x+y=n\\\!\mod N+1}}p(x,y)\op{x,y}{x,y}^{\msf{X}\msf{Y}},
\end{align}
which holds for all $n=0,\cdots,N$ and all $\phi\in[0,2\pi)$.  Here we are letting $\msf{X}$ and $\msf{Y}$ denote classical registers held by Alice and Bob, respectively, which store the classical outputs $x$ and $y$ of their local operations.  The distribution $p(x,y)$ is arbitrary, but the key constraint is that $p(x,y)=0$ whenever $x+y\not =n\mod N+1$ (which is why the sum appearing above is restricted).  The $\varphi^{A'B'}$ is some entangled resource state, and we would like to understand the amount of entanglement it must have for such a transformation to be possible.  

Recall that every LOSE instrument is a collection of CP maps $\{\mc{L}_{x,y}\}_{x,y}$, each of which is a convex combination of local CP maps, $\mc{L}_{x,y}=\sum_\lambda p(\lambda)\mc{A}^{AA'\to A}_{x|\lambda}\otimes\mc{B}^{BB'\to B}_{y|\lambda}$, and such that $\sum_x\mc{A}_{x|\lambda}$ and $\sum_y\mc{B}_{y|\lambda}$ are both trace-preserving for every $\lambda$.  This encompasses the most general strategy that Alice and Bob can employ without communicating with each other.  Since Eq. \eqref{Eq:LOSE-transform} describes a family of pure-state transformations, we do not need to consider mixtures generated by the random variable $\lambda$, and so without loss of generality we can assume that the LOSE instrument is a collection of product CP maps $\{\mc{L}_{x,y}=\mc{A}_{x}\otimes\mc{B}_y\}_{x,y}$. 

\begin{theorem}
\label{Thm:LOSE-entanglement}
The LOSE transformation in Eq. \eqref{Eq:LOSE-transform} is possible only if the entanglement entropy of the resource state satisfies $\msf{E}(\varphi^{A'B'}):=S(\varphi^{A'})\geq\log (N+1)$, where $S$ denotes the von Neumann entropy.
\end{theorem}

The theorem implies that winning the clock game requires the local ancilla states to have at least $D=N+1$ levels.  Furthermore, if both Alice and Bob have $D$-level systems locally available, they must be maximally entangled with each other to guarantee unit success in the game.  While Theorem \ref{Thm:LOSE-entanglement} is phrased in terms of the clock game, we stress that the clock game is an abstraction for any task in which the time-bin of an incident photon is learned by nondestructive local measurements.  In particular, the lower bound of $\log(N+1)$ corresponds with entanglement cost in the telescopy protocol of Ref. \cite{khabiboulline2019optical}, thereby proving its optimality.

\begin{remark}
While Eq. \eqref{Eq:LOSE-transform} is specified to hold for all choices of $\phi\in[0,2\pi)$, the same conclusion of Theorem \ref{Thm:LOSE-entanglement} holds if we just allow $\phi\in\{0,\pi\}$.  The proof below is carried out for this more restricted case.  
\end{remark}

\begin{proof}

Let us begin by taking operator-sum representations of the local maps,
\begin{align}
    \mc{A}_x(\cdot)&=\sum_{i}R_{x,i}(\cdot)R_{x,i}^\dagger,&\mc{B}_y(\cdot)&=\sum_{j}S_{y,j}(\cdot)S_{y,j}^\dagger.
\end{align}
To facilitate the pure-state transformations described by Eq. \eqref{Eq:LOSE-transform}, the Kraus operators must satisfy the condition
\begin{align}
\label{Eq:Kraus-transformation}
    R_{x,i}\otimes S_{y,j}\ket{\Psi_{\phi,n}}^{AB}\ket{\varphi}^{A'B'}=\gamma_{x,i,y,j|n,\phi}\ket{\Psi_{\phi,n}}^{AB}
\end{align}
for all $(x,i,y,j)$ and all $(n,\phi)$.  The coefficients $\gamma_{x,i,y,j|n,\phi}$ are complex numbers satisfying $\sum_{x,i,y,j}|\gamma_{x,i,y,j|n,\phi}|^2=1$ for all $(n,\phi)$.  We require that  $\gamma_{x,i,y,j|n,\phi}=0$ whenever $x+y\not=n\mod N+1$, which corresponds to the condition of Alice and Bob correctly identifying the time-bin $n$.  Since Eq. \eqref{Eq:Kraus-transformation} holds for every $\phi\in\{0,\pi\}$, by linearity we have
\begin{subequations}
\begin{align}
    R_{x,i}\otimes S_{y,j}\ket{0}^A\ket{n}^B\ket{\varphi}^{A'B'}&=\gamma_{x,i,y,j|n,\phi}\ket{0}^A\ket{n}^B\\
     R_{x,i}\otimes S_{y,j}\ket{n}^A\ket{0}^B\ket{\varphi}^{A'B'}&=\gamma_{x,i,y,j|n,\phi}\ket{n}^A\ket{0}^B.
\end{align}
\end{subequations}
Here we are using the short-hand notation $\ket{0}\equiv \ket{0,n}$ and $\ket{n}\equiv \ket{1,n}$.  For the bipartite state $\ket{0,n}^A\ket{1,n}^B$, we write $\ket{0;n}^{AB}\equiv \ket{0}^A\ket{n}^B$.  In terms of the CP maps $\mc{A}_x$ and $\mc{B}_y$, the previous equations take the form
\begin{subequations}
\begin{align}
    \mc{A}_x\otimes\mc{B}_y(\op{0;n}{0;n}^{AB}\otimes\varphi^{A'B'})&=p_{x,y|n}\op{0;n}{0;n}^{AB}\label{Eq:CP-cons-a}\\
    \mc{A}_x\otimes\mc{B}_y(\op{n;0}{n;0}^{AB}\otimes\varphi^{A'B'})&=p_{x,y|n}\op{n;0}{n;0}^{AB},\label{Eq:CP-cons-b}
\end{align}
\end{subequations}
where $p_{x,y|n}=\sum_{i,j}|\gamma_{x,i,y,j|n}|^2$.  In Eqns. \eqref{Eq:CP-cons-a} and \eqref{Eq:CP-cons-b}, let us take a trace of both sides and sum over $y$.  Since $\sum_{y}\mc{B}_y$ is trace-preserving, we have
\begin{align}
    \sum_yp_{x,y|n}&=\tr[\mc{A}_x(\op{0}{0}^A\otimes\varphi^{A'})]\notag\\
    &=\tr[\mc{A}_x(\op{n}{n}^A\otimes\varphi^{A'})],
\end{align}
which says that $q_x:=\tr[\mc{A}_x(\op{n}{n}^A\otimes\varphi^{A'})]$ forms a probability distribution that is independent of $n$.  Consequently, we can define density matrices for system $BB'$ given by
\begin{align}
    \sigma_y^{BB'}=\sum_{x=0}^{N}\tr_A[\mc{A}_{x}(\op{x+y}{x+y}^A\otimes\op{0}{0}^B\otimes\varphi^{A'B'})]\notag
\end{align}
for $y=0,\cdots,N$.  Here all addition is done modulo $N+1$.  But since $p_{x,y|n}=0$ if $x+y\not=n\mod N+1$, Eq. \eqref{Eq:CP-cons-b} implies that $\tr[\mc{B}_y(\sigma_{y'})]=\delta_{yy'}$.  Therefore, the $\sigma_y$ form a collection of $N+1$ mutually orthogonal states.  Hence the von Neumann entropy $S$ gives the bound \cite{Nielsen-2000a}
\begin{align}
    \log D\leq S\left(\frac{1}{N+1}\sum_{y=0}^{N}\sigma_y\right)=S(\varphi^{B'}),
\end{align}
since
\begin{align}
    \sum_{y=0}^{N}\sigma_y&=\sum_{x,y=0}^{N}\tr_A[\mc{A}_{x}(\op{y}{y}^A\otimes\op{0}{0}^B\otimes\varphi^{A'B'})]\notag\\
    &=\tr_{AA'}\left[\mathbb{1}^A\otimes\op{0}{0}^B\otimes\varphi^{A'B'}\right]=(N+1)\varphi^{B'}.\notag
\end{align}
This completes the proof.
\end{proof}

\subsection{Generalization to multiple parties} \label{sec:multi1}

We will now generalize the clock game so that it can involve $K\geq2$ parties. The updated rules are:
\begin{enumerate}
    \item The referee sends a phase-encoded state within the $n$-th time-bin
    \begin{equation} \label{eq:W_alpha}
    \begin{aligned}
        \ket{W_{\phi,n}} & =  \big(\ket{1,n}_1 \ket{0,n}_2...\ket{0,n}_K \\
        & + e^{i\phi_2} \ket{0,n}_1 \ket{1,n}_2...\ket{0,n}_K  \\
        & + ... \\
        & + e^{i\phi_K}\ket{0,n}_1 \ket{0,n}_2...\ket{1,n}_K\big) / \sqrt{K}
    \end{aligned}
    \end{equation}
    to $K$ parties, where $\ket{j,n}_k$ denotes $j$ excitations sent within the $n$-th time-bin to party $k$, and no excitations sent within the other time-bins. $n$ is the time-bin within which the parties received the excitation. The set of all possible time-bins $\{1,2,...,N\}$ is known to all parties. Only the referee knows $n$ and the phase shifts $\phi_i$.
    The referee is allowed to trick the parties by sending the vacuum state in all time-bins instead of the phase-encoded state. We reserve the index $n=0$ for such case and denote $\ket{W_{\phi,0}}$ as the corresponding vacuum state.
    \item The parties process the data sent from the referee along with some ancilla systems. The ancilla systems are allowed to be entangled states shared by the parties. The parties are free to specify the ancilla they want to receive, which allows for shared entanglement. The processing of available quantum states must be done locally, which can lead to modification of the quantum state received from the referee. We will denote to modified state as $\rho'$.
    
    \item The parties reply to the referee with the information $(x_1 , x_2, ..., x_K, \rho')$, where $x_i$'s are integers. 
    
    \item The referee measures the $\rho'$ state using the projective measurement $\{ \ket{W_{\phi,n}}\bra{W_{\phi,n}}, \mathbb{1} - \ket{W_{\phi,n}}\bra{W_{\phi,n}} \}$. The parties win the game if 
    \begin{equation}
        \sum_{i=1}^K x_i \ \text{mod } D = n
    \end{equation}
    and if the referee gets the outcome $\ket{W_\phi}$ in the measurement. 
\end{enumerate}

\subsection{Multi-party winning strategy} \label{sec:multiparty_singleQudit}

The procedure given in this chapter is a generalization of the game given in Sec. \ref{sec:pairwise_1pair}.

\begin{figure}
    \centering
    \includegraphics[width=8.6cm]{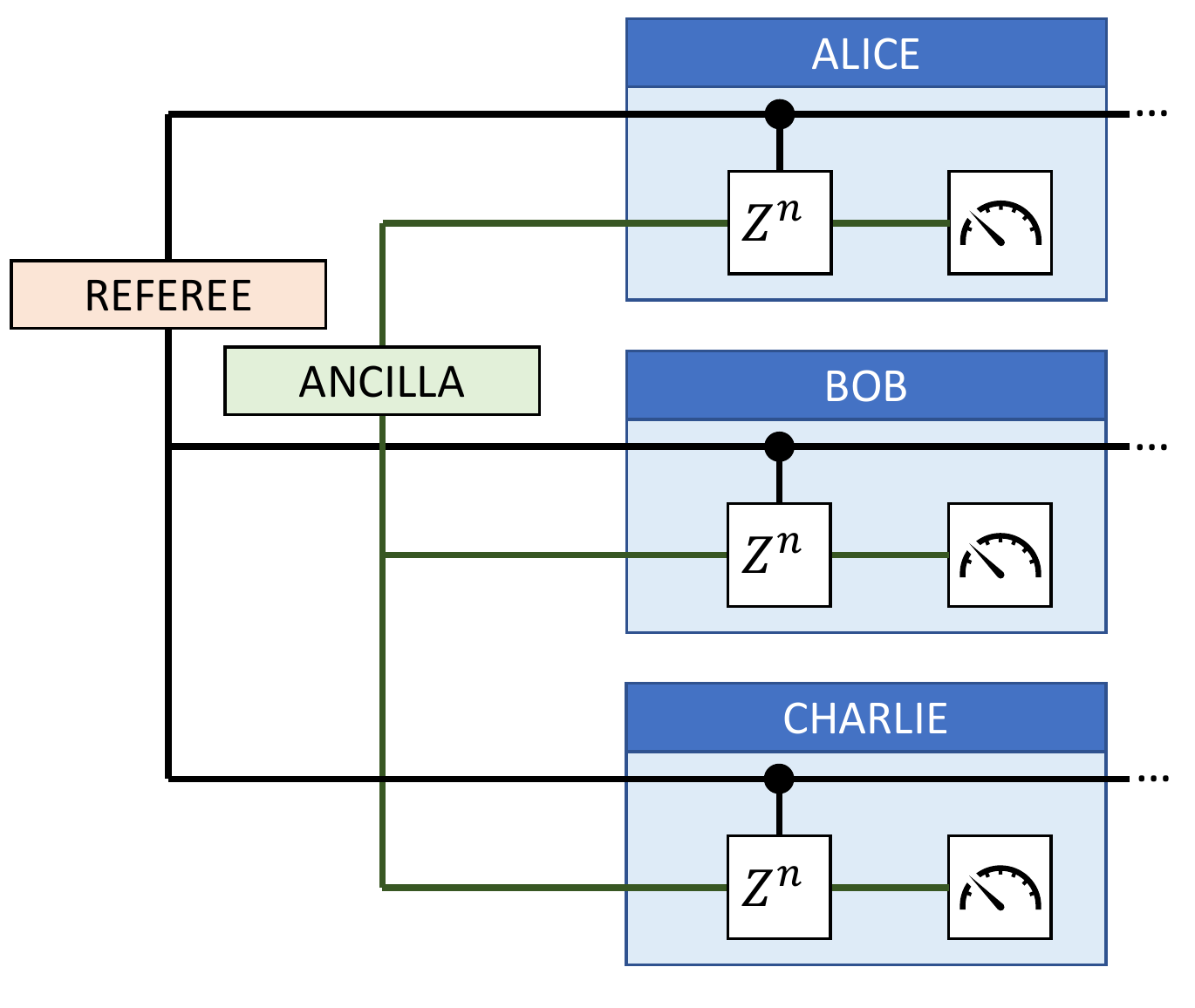}
    \caption{Circuit representation of the procedures performed by $K=3$ parties given that they have one entangled qudit state available. The measurements are performed in Fourier basis. Note that the procedure generalizes the scheme given in Fig. \ref{fig:pairwise_circuit}.}
    \label{fig:multiparty_scheme}
\end{figure}

We start the analysis by allowing the parties to share $K$ $D$-level systems prepared in the generalized GHZ state
\begin{equation} \label{eq:multiparty_ancilla}
\begin{aligned}
    \ket{\Phi_{D,0}^{(K)}} = \sum_{j=0}^{D-1} \ket{j}_1 \otimes \ket{j}_2 \otimes ... \otimes \ket{j}_K
\end{aligned}
\end{equation}
and requesting that the number of allowed time-bins  $N$ (time-bins) satisfies $N\leq D-1$. Note that the property (\ref{eq:Zn_effect}) generalizes to the state above: if one performs a $Z^n$ gate on any qudit in the state (\ref{eq:multiparty_ancilla}), then the resulting state is
\begin{equation}
\begin{aligned}
    \ket{\Phi_{D,0}^{(K)}} \rightarrow \ket{\Phi_{D,n}^{(K)}}
    & = \left[ Z^n \otimes \mathbb{1} \otimes ... \otimes \mathbb{1} \right] \ket{\Phi_{D,0}^{(K)}} \\
    & = \left[ \mathbb{1} \otimes Z^n \otimes ... \otimes \mathbb{1} \right] \ket{\Phi_{D,0}^{(K)}} \\
    & = \left[ \mathbb{1} \otimes \mathbb{1} \otimes ... \otimes Z^n \right] \ket{\Phi_{D,0}^{(K)}}    
\end{aligned}
\end{equation}
where $\ket{\Phi_n^{(K)}}$ can be expressed in the Fourier basis
\begin{equation} \label{eq:Phi_n_K}
\begin{aligned}
    \ket{\Phi_{D,n}^{(K)}} = \frac{1}{\sqrt{D}} & \sum_{\tilde{j_1}=0}^{D-1} \sum_{\tilde{j_2}=0}^{D-1} ... \sum_{\tilde{j_K}=0}^{D-1} \times \\ & \times \delta_{\tilde{j_1} + \tilde{j_2} + ... + \tilde{j_K},n}^{(D)} \ket{\tilde{j_1}} \ket{\tilde{j_2}} ... \ket{\tilde{j_K}}.    
\end{aligned}
\end{equation}

The procedures performed by the parties are summarized in Fig. \ref{fig:multiparty_scheme}. When the parties receive the referee qubits, they perform a $CZ^n$ gates with referee qubits as controls and ancilla qudits as targets. If the referee has supplied the W-state (\ref{eq:W_alpha}) in the $n$-th time bin, after the gates the ancilla state will be transformed to $\ket{\Phi_n^{(K)}}$. Next, the parties perform the measurements of the local ancilla qudits in the Fourier basis and obtain a set of results $\bar{j}_1, \bar{j}_2,...,\bar{j}_K$, with $\bar{j}_i$ being the result obtained by $i$-th party. All parties communicate their results to the referee and send back the quantum state they received from her. The referee computes
\begin{equation}
    n = \bar{j}_1 + \bar{j}_2 + ...  + \bar{j}_K \text{ mod }D.
\end{equation}
According to (\ref{eq:Phi_n_K}), this should return the time-bin within which she has provided the W-state, satisfying one of the winning conditions. The projective measurement she performs should return the right result, since the referee's quantum state remained unmodified after the local processing. 

\section{Application: Quantum-Enhanced Telescopy} \label{sec:clock_and_telescopes}

\begin{figure}
    \centering
    \includegraphics[width=8.6cm]{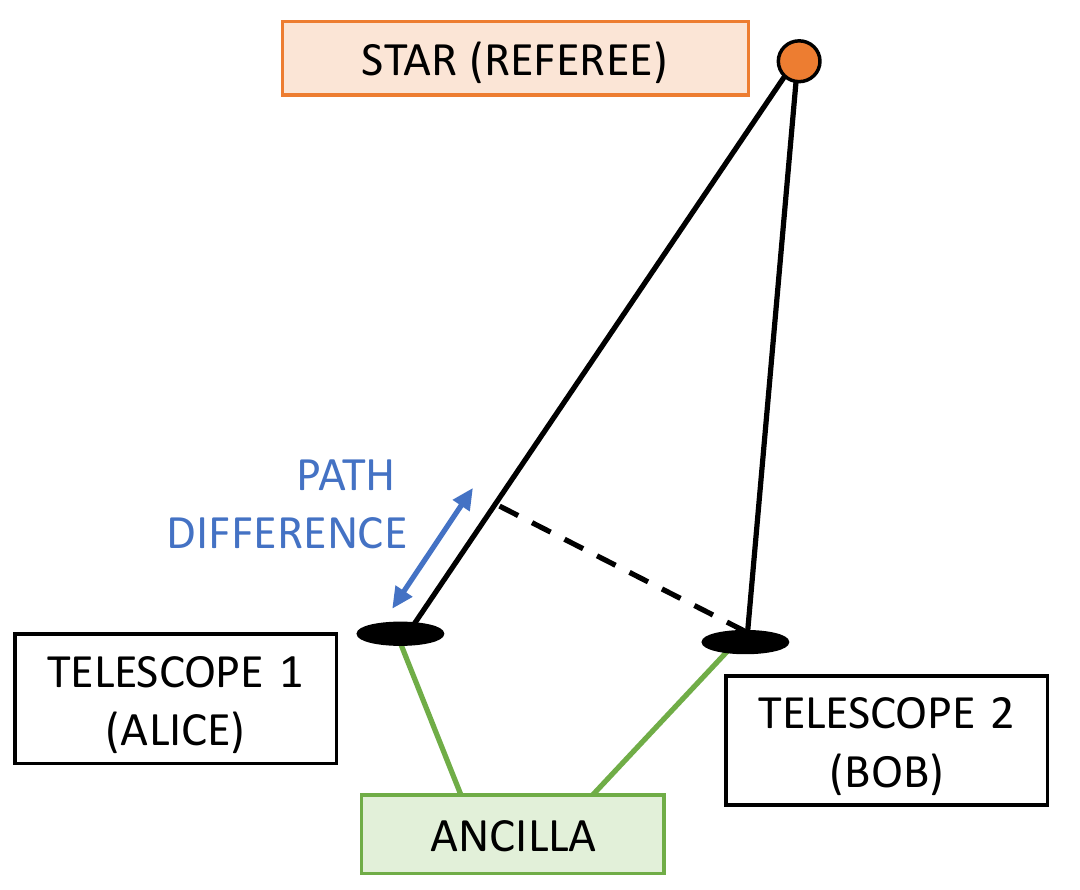}
    \caption{General scheme of quantum-enhanced long-baseline interferometry. The blue color indicates the path difference, which gives rise to the relative phase shift.}
    \label{fig:analogy}
\end{figure}

An interesting application of the games given above is determining the photon arrival time-bin in quantum-enhanced long-baseline telescopy. Consider a stellar source that supplies radiation to a pair of telescopes held by Alice ($A$) and Bob ($B$), respectively, as shown in Fig.~\ref{fig:analogy}. We assume that the source can be described by a weak thermal state. For a given time-bin $i$, the state of incoming radiation has the form \cite{tsang2011quantum}
\begin{equation} \label{eq:single_bin_state}
    \rho_i = (1-\epsilon_1)\rho_{0,i} + \epsilon_1 \rho_{1,i} + \mathcal{O}(\epsilon_1^2),
\end{equation}
where 
\begin{equation} \label{eq:2telescopes_1photon}
\begin{aligned}
    \rho_{0,i} & = \op{0}{0}^{A}\otimes\op{0}{0}^{B}\\
    \rho_{1,i} & = \frac{1}{2}\big( \op{i}{i}^{A}\otimes\op{0}{0}^{B} + \op{0}{0}^{A}\otimes\op{i}{i}^{B} \\
        & +  \nu^* \op{0}{i}^{A}\otimes\op{i}{0}^{B} + \nu \op{i}{0}^{A}\otimes\op{0}{i}^{B} \big)
\end{aligned}
\end{equation}
describe the time-bins in which the star supplies zero ($\rho_{0,i}$) and one photon ($\rho_{1,i}$) to the telescopes. Note, here we are adopting the notation from Sec. \ref{sec:LOSE} that $\ket{0}=\ket{0,i}$ and $\ket{i}=\ket{1,i}$. The $\mathcal{O}(\epsilon_1^2)$ term in Eq. \eqref{eq:single_bin_state} describes two or more photon events and is assumed to be negligible.  The goal of the procedure is to determine the complex visibility $\nu$. The visibility is a function of the baseline connecting the telescopes, it can be used to compute the intensity profile of the examined stellar source using the van Cittert-Zernike theorem \cite{van_Cittert_Zernike, van_Cittert_Zernike_1}.  

One way to estimate $\nu$ it is to physically bring the light from the two telescopes together.  However, this so-called direct detection method suffers from losses that occur when we try to transfer the stellar photons from one location to the other.  Another approach is to perform all measurements locally. However, it was shown in Ref. \cite{tsang2011quantum} that this performs significantly worse than the direct detection method. A clever work-around was proposed by Gottesman \textit{et al.} that uses local measurements and quantum teleportation to simulate direct detection \cite{gottesman2012longer}. Their scheme includes distributing single photons to the telescope locations (ancilla with shared entanglement) and interfering them using beam splitters with the stellar photons. One measures the output ports of the beam splitters in the photon-number basis, and coincidence counts provide information about the visibility. A serious drawback of this scheme is that it requires an extremely high entanglement generation rate. In principle, one wants to perform measurements on as many stellar photons as possible, and the teleportation-based protocol requires distributing one entangled ancilla state within each available time-bin. With current technology, this task is not feasible. 


A significant improvement can be made to this protocol if the time-bin of the incident photon can first be ascertained before performing the visibility measurements on the occupied spatio-temporal mode \cite{khabiboulline2019optical}.  In more detail, consider stellar radiation in the weak thermal light regime ($\epsilon_1 \ll 1$) that arrives at the telescopes within $N$ time-bins.  
We assume that for each time-bin the incoming stellar photon state is described by (\ref{eq:single_bin_state}) and the photonic states within each time bin are independent of each other. Note that the probability of the photon arriving to one of the telescopes in time-bin $i$ can be considered as a Bernoulli trial with the success probability of $\epsilon_1$, and the probability of $k$ photons arriving within $N$ time-bins is described by a Bernoulli distribution
\begin{equation}
    P(k;N) = {{N}\choose{k}} k^{\epsilon_1} (N-k)^{\epsilon_1}.
\end{equation}
The probability that exactly one stellar photon will arrive within $N$ time-bins is
\begin{equation}
    \epsilon \equiv P(1;N) = N \epsilon_1 (1-\epsilon_1)^{N-1}.
\end{equation}
In the regime where we expect at most one stellar photon to arrive within $N$ time bins, the state of the incoming radiation can be described as
\begin{equation} \label{eq:multibin_stellar_photon}
    \rho = (1-\epsilon)\rho_0 + \frac{\epsilon}{N}\sum_{i=1}^N \rho_{1,i} \bigotimes_{j=1}^{N} {}^{'} \rho_{0,j}
\end{equation}
where the primed tensor product indicates that we include all the terms except $j = i$. The first term denotes no photons arriving at the telescopes across all the time-bins. The terms in the second sum describe one photon arriving within time-bin $i$ and no photons arriving within the other time-bins, with $\rho_{0,i}$ and $\rho_{1,i}$ given by (\ref{eq:2telescopes_1photon}).

Suppose now that one is able to perform a quantum nondemolition (QND) measurement that post-selects on one of the terms within the sum in (\ref{eq:multibin_stellar_photon}). Such a measurement corresponds to determining whether or not the stellar photon has arrived and, if it has, determining the arrival time-bin.  Crucially, this measurement needs to be done without destroying the information about the visibility.  If such a QND measurement were performed, it would greatly simplify the task of determining the visibility since it would allow one to work with the state $\rho_{1,i}$ defined in (\ref{eq:2telescopes_1photon}) instead of (\ref{eq:multibin_stellar_photon}), which is heavily dominated by the vacuum. The protocol of Gottesman \textit{et al.} could then be directly performed on $\rho_{1,i}$.

We observe that the necessary QND measurement can be achieved by performing the winning strategy in the two-party clock game described in Sec. \ref{sec:pairwise_1pair}. In the telescopy setup, the stellar source plays the role of the referee, and the separated quantum telescopes play the role of Alice and Bob (Fig. \ref{fig:analogy}). The task is to determine when the star (referee) has supplied the photon. Note that even though the state of the stellar photon within the occupied time-bin is not described by a pure state (\ref{eq:ref_bell_pair}) but by a density matrix $\rho_{1,i}$, the scheme of the photon arrival time-bin measurement remains unchanged. 

\begin{figure}
    \centering
    \includegraphics[width=8.6cm]{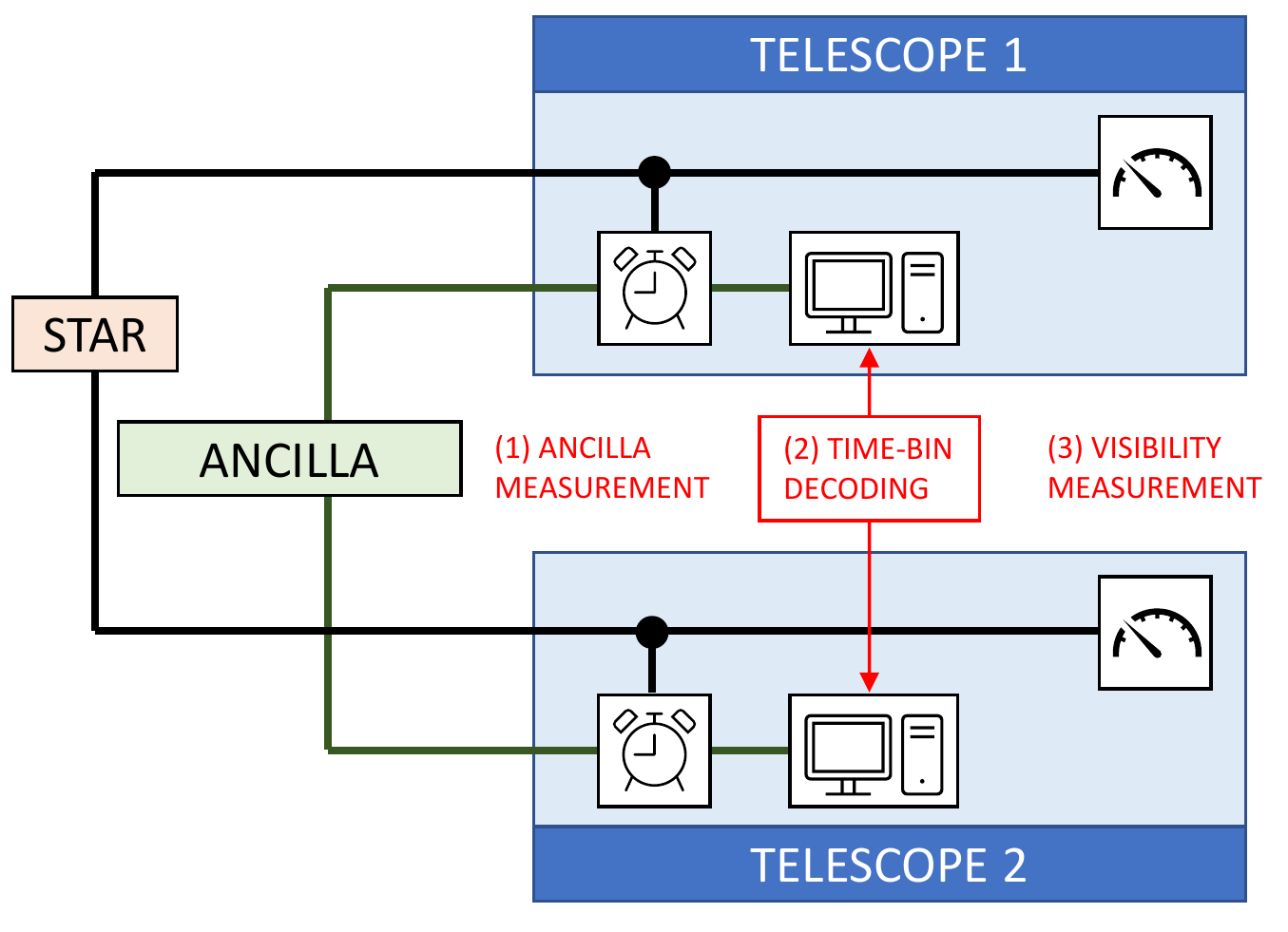}
    \caption{Circuit diagram of quantum-enhanced long-baseline interferometry with the photon arrival time-bin measurement.}
    \label{fig:telescope_scheme}
\end{figure}

This is because the clock game works for any phase shift within the phase-encoded state. Suppose that the source to be examined is a set of point sources indexed by $q$. If the source $q$ emits a photon, it will have to follow a different path to reach both telescopes (see Fig. \ref{fig:telescope_scheme}); the path difference gives rise to a relative phase shift $\phi_q$. Observe that if the stellar photon was supplied in time-bin $i$ by source $q$, then the state of spatio-temporal modes reaching the telescopes is described by the phase-encoded state $\ket{\Psi_{{\phi_q},i}}$ defined in Eq. (\ref{eq:ref_bell_pair}). However, we cannot be certain about which source provided the photon. Let $p_q$ denote the probability that it was source $q$ that provided the photon given that the photon has arrived from the sources. Then the incoming state given that the stellar photon has arrived in time-bin $i$ is
\begin{equation}
    \tilde{\rho}_{1,i} = \sum_q p_q \ket{\Psi_{{\phi_q},i}} \bra{\Psi_{{\phi_q},i}}.
\end{equation}
If one defines $\nu=\sum_q p_q \exp(-i\phi_q)$, then the state above agrees with (\ref{eq:2telescopes_1photon}). For extended sources one would replace the sum over $q$ by an integral. 

We observe that the states provided by the referee in the clock game by a weak stellar source in long-baseline interferometry become similar if we allow the referee to randomize the phase. In that case, different phases chosen by the referee correspond to different stellar point sources emitting the photon. However, the clock game works for any phase. Therefore, it can be applied in long-baseline interferometry to determine the stellar photon arrival time-bin. 

The only difference in the protocol is that after the measurements of the local ancilla, the parties communicate the results to each other and both of them determine the time-bin. Once they know when the stellar photon has arrived, they perform the visibility measurement on the appropriate spatio-temporal time-bin. 

Sections \ref{sec:entanglement} and \ref{sec:classical_communication} provide us with the requirements that a time-bin measurement should satisfy. Such measurement is impossible if one does not distribute sufficient entanglement to the telescope locations and if one does not allow classical communication between them.  As shown in Fig. \ref{fig:telescope_scheme}, the decoding of the time-bin is followed by the visibility measurement on the occupied spatio-temporal time-bins. If the visibility measurement is complicated enough so that it can be performed only once across all the time-bins, then both parties need to know the occupied time-bin in advance so that they know when to perform the visibility measurement. 
The time-bin decoding requires classical communication between the parties, and the state of all arriving stellar modes must be stored until both parties finish the communication between them and determine the time-bin. The necessity of quantum memory can be avoided if one is able to perform fast visibility measurement for all time-bins. Then, after the time-bin is decoded, both parties select the result corresponding to the occupied mode and use it for visibility estimation. 

The idea of determining the stellar photon arrival time-bin before the visibility measurement was first applied by Khabiboulline \textit{et al.} \cite{khabiboulline2019optical}. Their scheme includes encoding both the state of the stellar photon and the time-bin on a set of auxiliary qubits, and then encoding the time-bin on another set of qubits. After that, the time-bin is decoded from the second set of qubits while leaving the visibility encoded in the first set. 

The strategy introduced in Sec. \ref{sec:pairwise_1pair} provides a more general approach to time-bin measurement because it allows the ancilla to have an arbitrary number of levels. One can apply it when the local ancilla states are $n_D$ qudits with $D$ levels, since they can be treated as $D^{n_D}$ level systems. If one has $2n_2$ ancilla qubits, then one can examine $2^{n_2}-1$ time-bins. However, recently for certain type of quantum systems, the third level has been explored (e.g., transmon qubits \cite{shlyakhov2018quantum}), so that they can be used as qutrits. Then, with the same amount of ancilla systems one can explore $3^{n_3}-1$ time-bins, where $n_3$ is the number of qutrits. The winning strategy described in this manuscript also applies to this case.

The multipartite version of the clock game can be applied to the setups involving more than two telescopes in distant locations. Suppose that our setup involves $M$ telescopes. In such a case, the incoming state from the stellar source can still be described by equation in the form of (\ref{eq:multibin_stellar_photon}), but now the vacuum term pertains to all the telescopes. The $\rho_{1,i}$ term  describes an entangled state of a single photon coherently arriving to the set of $M$ telescopes; its nonzero matrix elements are
\begin{equation}
\begin{aligned}
    & {}^{A_\alpha}\bra{i}\rho_{1,i}\ket{i}^{A_\alpha} = 1/N, \\ 
    & {}^{A_\alpha}\bra{i}\rho_{1,i}\ket{i}^{A_\beta} = \nu_{\alpha\beta}/N, \\ 
    & \nu_{\alpha\beta}=\nu_{\beta\alpha}^*, 
\end{aligned}
\end{equation}
where $A_\alpha$ and $A_\beta$ label telescopes $\alpha,\beta\in\{1,2,\cdots M\}$, the state $\ket{i}^{A_\alpha}$ describes one photon arriving at telescope $A_\alpha$ and no photons arriving at the other telescopes, and $\nu_{\alpha\beta}$ is the visibility associated with the baseline connecting the telescopes $A_\alpha$ and $A_\beta$. As in the two-telescope case, it can be advantageous to post-select the time-bins within which the stellar photon has arrived, and the multipartite version of the clock game can be used to do so. As before, by linearity and the fact that the clock game holds for all phases $\phi$, the procedure is still valid despite the fact that we do not work with pure states (\ref{eq:W_alpha}). 

The multi-party clock game can be used as a subroutine in visibility measurements that involve multiple telescopes. An example is the scheme of Gottesman \textit{et al.} \cite{gottesman2012longer} where a single photon needs to be distributed to the set of telescopes in a W state for each time-bin one expects the stellar photon to arrive. As in the two-telescope case, inclusion of the clock game allows one to determine the time-bin prior to the visibility measurement at a lower entanglement cost than in the original protocol. For $N$ possible time-bins and a set of $M$ telescopes, the Gottesman \textit{et al.} protocol requires the distribution of $N$ W states made out of $M$ qubits. The protocol supported by the clock game would require one such state for visibility measurement and one entangled state of $M$ qudits with $N+1$ levels for the time-bin estimation performed prior to the visibility measurement in the clock game subroutine. 

To make this comparison more clear, assume that one uses five telescopes to examine, say, $N=1,\!023$ time-bins and one can use only qubits. The scheme of Gottesman \textit{et al.} requires $1,\!023$ entangled states of five qubits, consuming $5,\!115$ qubits. The scheme supported by the clock game requires one entangled state of five qubits for visibility measurement and a set of 50 entangled qubits for the clock game. The 50 qubits are distributed equally between five telescope locations with each party receiving a set of 10 qubits (note that it forms a $1,\!024$-level qudit required for the clock game). 
Therefore, the scheme supported by the clock game consumes $55$ qubits, significantly less than $5,\!115$ qubits in the unsupported protocol. We note that a similar improvement was achieved in Ref. \cite{Khabiboulline-2019b}, but the clock game achieves the following advantages: (1) it allows to use qudits instead of just qubits in time-bin estimation, (2) it isolates the task of time-bin estimation so that it can be used for other visibility measurement schemes, and (3) it abstracts the task of time-bin estimation to a task formulated within the framework of quantum information so that it can be used in other fields.

\section{Phase extraction protocol}

\label{Sect:linear-phase-extraction}

As described in the previous section, once the time-bin of the stellar photon is acquired, the star's visibility $\nu$ with respect to the two telescopes can be determined using the original scheme by Gottesman \textit{et al.} \cite{gottesman2012longer}. Apart from the distribution of entanglement between the two telescopes, the latter protocol just involves a local phase shifter and beam splitters, combined with classical post-processing.  

In any telescopy protocol, the amount of information gathered about the visibility per stellar photon can be quantified using the Fisher information. When optimized over all (unrestricted) quantum measurements, one obtains the quantum Fisher information \cite{Paris-2009a}.  The quantum Cr\'{a}mer-Rao bound says that the inverse of the Fisher information lower bounds the variance of any unbiased estimator for the unknown parameter.  However, in this case the visibility $\nu$ is a complex value, consisting of two unknown parameters (its real and imaginary parts).  Hence, in general one is left with a multi-parameter estimation problem in which the Cr\'{a}mer-Rao bound is replaced by matrix inequalities \cite{Yang-2019a}. 

To simplify the discussion going forward, let us assume that, after being detected in a known time-bin, the stellar photon is in a pure state of the form 
\begin{equation}
    \label{Eq:state-QFI}
    \ket{\Psi_s}=\frac{1}{\sqrt{2}}(\ket{0}\ket{1}+e^{i\phi}\ket{1}\ket{0}).
\end{equation}
In this case, the visibility is simply $\nu=e^{-i\phi}$ and $\phi$ is a single parameter to be estimated.  The protocol of Gottesman \textit{et al.} attains a Fisher information of $\frac{1}{2}$ due to the fact that the two-photon interference measurement yields no information half of the time.  
Here we describe a protocol in which the Fisher information can be made arbitrarily close to one using only linear optical elements and shared entanglement.  More precisely, each telescope needs to only perform unitaries that locally preserve the photon number.  The trade-off, however, is that more and more entanglement is needed to be shared between the telescopes to drive the Fisher information closer and closer to one.  While we describe the protocol below in terms of estimating the single parameter $\phi$, we remark that the protocol also works in the general case of estimating an arbitrary complex visibility $\nu$, and it consumes half as many stellar photons compared to the Gottesman \textit{et al.} scheme.

Let Alice control the left (L) telescope and Bob control the right (R) one.  Suppose that $n$ ancilla states are distributed to them, each of the form $\frac{1}{\sqrt{2}}(\ket{0}+e^{i\delta}\ket{1})$.  We can write the full ancilla state as
\begin{align}
    \ket{\Psi_a}&=\frac{1}{2^{n/2}}\bigotimes_{i=1}^{n}(\ket{01}+e^{i\delta}\ket{10})_{2i,2i+1}\notag\\
    &=\frac{1}{2^{n/2}}\sum_{k=0}^ne^{i\delta k}\sum_{\Vert\mbf{x}\Vert=k}\ket{\mbf{x}}_{L_a}\otimes\ket{\ol{\mbf{x}}}_{R_a},
\end{align}
where $\mbf{x}$ is an $n$-bit string with Hamming weight $\Vert\mbf{x}\Vert$, and $\ol{\mbf{x}}$ denotes its bitwise complement.  Consider a photon emitted from a point source that reaches the telescopes telescopes in state $\ket{\Psi_s}$ with $\phi$ being an unknown phase.  The total $(n+1)$-photon state is given by (up to a normalization factor)
\begin{align}
\label{Eq:Full-state}
    &\ket{\Psi_s}\ket{\Psi_a}\notag\\
    &=\sum_{k=0}^{n-1}e^{i\delta k}\bigg(e^{i\delta }\sum_{\Vert\mbf{x}\Vert=k+1}\ket{0}_{L_s}\ket{\mbf{x}}_{L_a}\otimes\ket{1}_{R_s}\ket{\ol{\mbf{x}}}_{R_a}\notag\\
    &\qquad +e^{i\phi}\sum_{\Vert\mbf{x}\Vert=k}\ket{1}_{L_s}\ket{\mbf{x}}_{L_a}\otimes\ket{0}_{R_s}\ket{\ol{\mbf{x}}}_{R_a}\bigg)\notag\\
    &\qquad+\ket{0}_{L_s}\ket{0\cdots 0}_{L_a}\otimes\ket{1}_{R_s}\ket{1\cdots 1}_{R_a}\notag\\
    &\qquad+e^{i\delta(n+\phi)}\ket{1}_{L_a}\ket{1\cdots 1}_{L_s}\otimes\ket{0}_{R_a}\ket{0\cdots 0}_{R_s}.
\end{align}
Notice that each term here has $k+1$ particles localized at the left telescope and $n-k$ particles localized at the right telescope.  It will be helpful to relabel the terms in parentheses as
\begin{align}
    &e^{i\delta}\sum_{j=0}^{\binom{n}{k+1}-1}\ket{j,k+1}_{L}\otimes\ket{j,n-k}_{R}\notag\\
    +&e^{i\phi}\sum_{j=\binom{n}{k+1}}^{\binom{n+1}{k+1}-1}\ket{j,k+1}_{L}\otimes\ket{j,n-k}_{R},
\end{align}
where $j$ is an index over all the states with $k+1$ particles on Alice's side (one quanta per mode) and $n-k$ particles on Bob's.

\begin{figure}
    \centering
    \includegraphics[width=8.6cm]{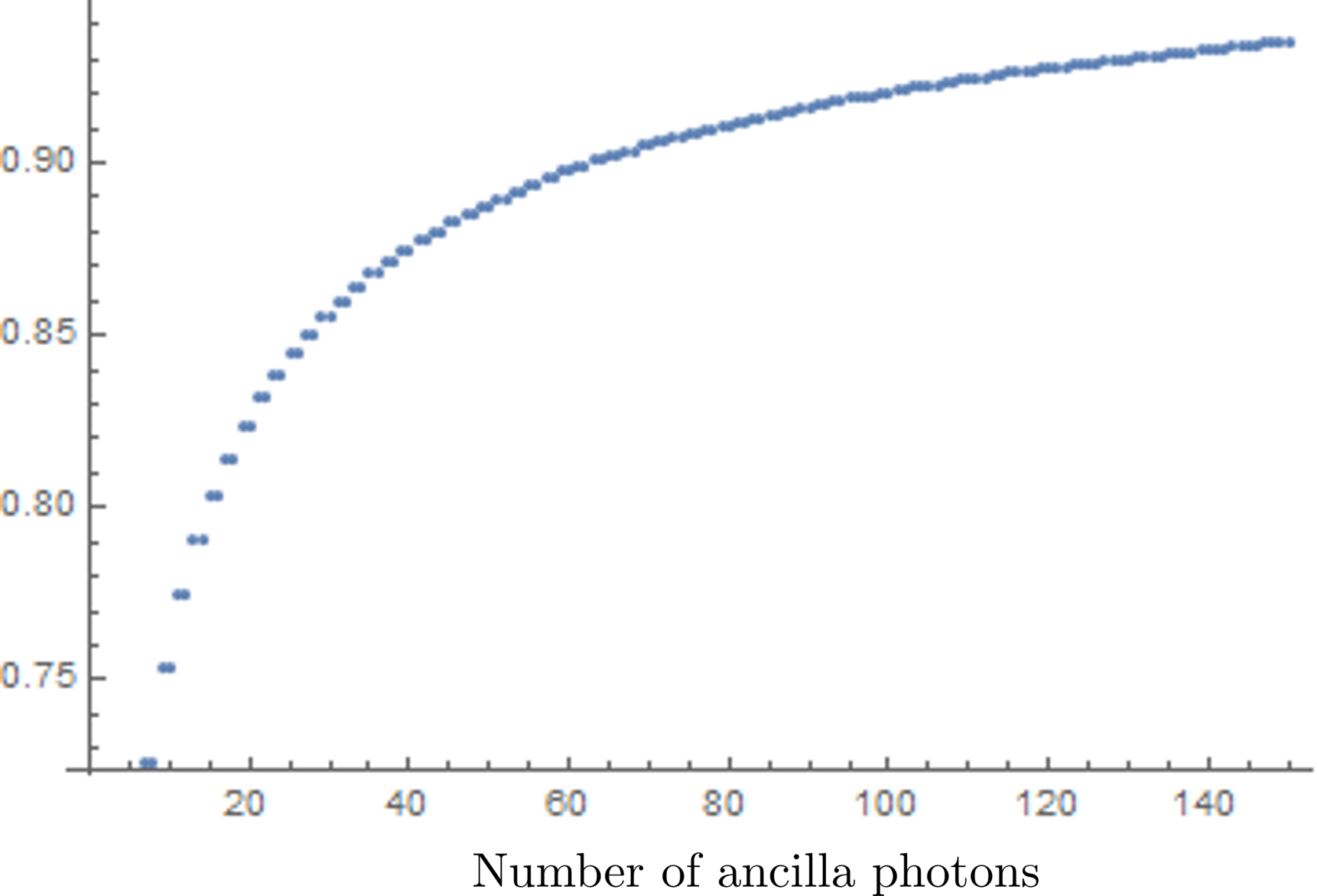}
    \caption{The average Fisher information per ancilla photon ($y$ axis) as a function of ancilla photon number ($x$ axis).  The phase angle $\Phi$ is sampled uniformly over the interval $[0,2\pi)$.}
    \label{Fig:Fisher_Information}
\end{figure}

In the first stage of the protocol, Alice performs a Fourier transformation on each block of $k+1$ particles for $k=0,\cdots,n-1$.  Each term in the large parentheses of Eq. \eqref{Eq:Full-state} will transform to
\begin{align}
    &e^{i\delta}\sum_{j=0}^{\binom{n}{k+1}-1}\ket{\wt{j,k+1}}_{L}\otimes\ket{j,n-k}_{R}\notag\\
    +&e^{i\phi}\sum_{j=\binom{n}{k+1}}^{\binom{n+1}{k+1}-1}\ket{\wt{j,k+1}}_{L}\otimes\ket{j,n-k}_{R},
\end{align}
Alice then measures each of her $n+1$ subsystems and tells Bob which ones of them contained a photon.  If none of them contain a photon or all of them do, then they abort (these correspond to the last two lines in Eq. \eqref{Eq:Full-state}.  On the other hand, if Alice detects $k+1$ photons for $k=0,\cdots,n-1$, then Alice tells Bob the particular configuration of clicks, which is labeled by some integer $j'\in\{0,1,\cdots,\binom{n+1}{k+1}-1\}$.  Bob's post-measurement state will be a superposition of the $\ket{j,n-k}_R$ with relative phases depending on the particular value of $j'$. Bob can correct these phases by controlled-phase gates, and his post-measurement state will be given by
\begin{align}
    e^{i\delta}\sum_{j=0}^{\binom{n}{k+1}-1}\ket{j,n-k}_{R}+e^{i\phi}\sum_{j=\binom{n}{k+1}}^{\binom{n+1}{k+1}-1}\ket{j,n-k}_{R}.
\end{align}
This can be expressed in normalized form as
\begin{align}
    \sqrt{\frac{n-k}{n+1}}\ket{0'}+\sqrt{\frac{k+1}{n+1}} e^{i(\phi-\delta)}\ket{1'},
\end{align}
where 
\begin{align}
    \ket{0'}&=\frac{1}{\sqrt{\binom{n}{k+1}}}\sum_{j=0}^{\binom{n}{k+1}-1}\ket{j,n-k}_{R},\notag\\
    \ket{1'}&=\frac{1}{\sqrt{\binom{n}{k}}}\sum_{j=\binom{n}{k+1}}^{\binom{n+1}{k+1}-1}\ket{j,n-k}_{R}.
\end{align}
The key point is that Bob's system has now collapsed into a two-dimensional subspace spanned by two orthogonal states $\{\ket{0'},\ket{1'}\}$.  He then rotates $\ket{0'}\mapsto\sqrt{1/2}(\ket{0'}+\ket{1'})$, $\ket{1'}\mapsto\sqrt{1/2}(\ket{0'}-\ket{1'})$ and then measures.  The outcome probabilities are given by
\begin{align}
    p(0'|k+1)&=\frac{1}{2}\left|\sqrt{\frac{n-k}{n+1}}+\sqrt{\frac{k+1}{n+1}} e^{i(\phi-\delta)}\right|^2\notag\\
    &=\frac{1}{2}\left(1+\frac{2\sqrt{(n-k)(k+1)}}{n+1}\cos(\phi-\delta)\right)\notag\\
    p(1'|k+1)&=\frac{1}{2}\left|\sqrt{\frac{n-k}{n+1}}-\sqrt{\frac{k+1}{n+1}} e^{i(\phi-\delta)}\right|^2\notag\\
    &=\frac{1}{2}\left(1-\frac{2\sqrt{(n-k)(k+1)}}{n+1}\cos(\phi-\delta)\right).
\end{align}
We are interested in computing the Fisher information of this protocol.  Note that
\begin{align}
\label{Eq:probabilities}
    \sum_{i=0}^1\tfrac{1}{p(i'|k+1)}\left[\tfrac{\partial p(i'|k+1)}{\partial\phi}\right]^2
    =\frac{\sin^2(\phi-\delta)}{\frac{(n+1)^2}{4(n-k)(k+1)}-\cos^2(\phi-\delta)}.
\end{align}
Hence the Fisher information is given by
\begin{align}
\label{Eq:Fisher-1}
    &\sum_{k=0}^{n-1}\text{Pr}(k+1)\frac{\sin^2(\phi-\delta)}{\frac{(n+1)^2}{4(n-k)(k+1)}-\cos^2(\phi-\delta)}\notag\\
    &\quad=\frac{1}{2^{n+1}}\sum_{k=0}^{n-1}\binom{n+1}{k+1}\frac{\sin^2(\phi-\delta)}{\frac{(n+1)^2}{4(n-k)(k+1)}-\cos^2(\phi-\delta)},
\end{align}
where $\text{Pr}(k+1)$ is the probability that $k+1$ particles are detected when measuring on the left telescope.  To put a lower bound on \eqref{Eq:Fisher-1}, we use a typicality argument.  Since the expected number of particles detected is $(n+1)/2$, let us say that a value $k+1$ is $\epsilon$-typical if $|k-(n+1)/2|<\epsilon(n+1)$, where $\epsilon>0$ is arbitrarily small.  Then the Fisher information is no less than
\begin{align}
    &\!\!\!\!\sum_{\text{$\epsilon$-typical $k+1$}}\text{Pr}(k+1)\frac{\sin^2(\phi-\delta)}{\frac{(n+1)^2}{4(n-k)(k+1)}-\cos^2(\phi-\delta)}\notag\\
    &\geq \text{Pr}(\text{$\epsilon$-typical $k+1$})\frac{\sin^2(\phi-\delta)}{\frac{(n+1)^2}{\left(n(1-2\epsilon)-1-2\epsilon\right)^2}-\cos^2(\phi-\delta)}.
\end{align}
However, as $n\to\infty$ we have $\text{Pr}(\text{$\epsilon$-typical $k+1$})\to 1$ and $\frac{(n+1)^2}{\left(n(1-2\epsilon)-1-2\epsilon\right)^2}\to 1+O(\epsilon)$.  This implies that the Fisher information can be made arbitrarily close to $1$, which is optimal for phase measurements.  Hence we have established the following result.
\begin{proposition}
For stellar point sources [i.e., states having the form of Eq. \eqref{Eq:state-QFI}], the quantum Fisher information for parameter $\phi$ can be attained by using local linear optical operations and shared entanglement (see Fig. \ref{Fig:Fisher_Information}).
\end{proposition}

The protocol presented in this section becomes more practical if prior to performing it one performs the clock game procedure. High values of Fisher information are achieved with high values of entangled pairs (see Fig. \ref{Fig:Fisher_Information}), e.g.,  achieving the Fisher information of 0.85 requires about 30 entangled pairs. Distributing such number of entangled pairs within each time-bin one expects a stellar photon can become impractical since most of the time-bins are not occupied. A possible solution is to use the clock game to determine when the stellar photon has arrived and apply the phase extraction for that time-bin. 

\section{Conclusions} \label{sec:conclusions}

In this manuscript, we have considered the clock game formulated in the framework of quantum information theory, which can be applied as a subroutine in quantum-enhanced long-baseline interferometry. The winning strategy provides a method for quantum nondemolition measurement of the photon arrival time-bin. We have considered the resources required to win the game in terms of the necessary degree of entanglement within it, and we have shown our winning strategy of the clock game achieves the task with the least possible resources. Notably, we proved that $\log (N+1)$ shared ebits is needed to discriminate between $N$ time-bins without disturbing the relative phase between laboratories, which matches the upper bound of Ref. \cite{khabiboulline2019optical} and the winning strategy for the clock game introduced here.  Errors introduced to the ancilla state lead to a decrease in the probability of winning the game that we have quantified in the case of amplitude damping and dephasing. 

Later, we have examined the task of the phase extraction within an entangled state with the restriction that the local operations must be linear, i.e., must conserve the local number of excitations. Our scheme provides an optimal measurement of the phase in the sense that it achieves the maximum allowed value of the Fisher information. However, improving the Fisher information requires increasing the number of ancillary qubits. 

Our schemes can be used as elements of other quantum-enhanced telescopy procedures. The winning strategy of the clock game provides a protocol for learning the stellar photon's arrival time-bin, but it does not depend on the type of measurement that is used to extract the information about the visibility. Therefore, one can use it to verify which spatio-temporal modes are occupied by stellar photons, and then perform the preferred method of the visibility measurement. 

It should be noted that the implementation of the clock game in practical setups might require additional research related to the context within which the clock game scheme is implemented. For example, implementing the clock game in long-baseline interferometry requires figuring out the optimal dimensionality of the qudits.

\begin{acknowledgments}
We acknowledge Rayleigh W. Parker for the conversations that stimulated the development of this paper, and Paul G. Kwiat for his valuable comments.
We thank Virginia O. Lorenz, John D. Monnier, Michael G. Raymer, and Brian J. Smith for helpful discussions. This work was supported by the multi-university National Science Foundation Grant No. 193632 - QII-TAQS: Quantum-Enhanced Telescopy.
\end{acknowledgments}

\bibliography{apssamp}

\newpage

\appendix

\section{Necessity of Entangled Ancilla in the Clock Game} \label{sec:entanglement}

In Sec. \ref{sec:LOSE} we have examined the resources needed to win the pairwise clock game based on the framework of LOSE transformations. In the Appendixes \ref{sec:entanglement}-\ref{sec:ent_dim} we consider various limitations introduced to the ancilla and see how they affect the possibility of winning the clock game. The ancilla will be treated as a meter used to measure the phase-encoded state arrival time-bin.

In this section, we examine the possibility of winning the clock game described in Sec. \ref{sec:clock_game} in a local way, i.e., we will not allow an entangled ancilla and exchange of quantum information between the parties. 

First, we note that Alice and Bob have restricted knowledge about the state sent by the referee. That lack of knowledge can be formulated mathematically by representing the state they receive not by a pure state (\ref{eq:ref_bell_pair}), but by a mixed state
\begin{equation} \label{eq:encoded_state_2}
    \rho_r = \int_{\phi_0}^{\phi_0+2\pi}d\phi \sum_{n=0}^N p(n,\phi) \ket{\Psi_{\phi,n}} \bra{\Psi_{\phi,n}} 
\end{equation}
with $\ket{\Psi_{\phi,n}}$ defined in (\ref{eq:ref_bell_pair}) and (\ref{eq:n0}). For $n>1$, $p(n,\phi)$ is the probability that Alice and Bob will receive the excitation within the $n$-th time-bin with the encoded phase $\phi$; $p(0,\phi)$ is the probability that the referee does not sent the phase-encoded state. $\phi_0$ is the arbitrary reference angle. Note that the state $\ket{\Psi_{\phi,0}}$ is the vacuum state and does not have the phase encoded in it. To keep (\ref{eq:encoded_state_2}) valid, we can assign that vacuum state to some angle, e.g. $\phi_0$, and make $p(0,\phi) = p(n=0)\delta(\phi-\phi_0)$. 

Since (\ref{eq:encoded_state_2}) represents the knowledge Alice and Bob have about the received state, $p(n,\phi)$ represents their degree of belief and can depend on the nature of the problem. 

The main task in the game is to measure the time-bin within which the referee has sent the excitation. It can be achieved by coupling $\rho_r$ to the meter (ancilla), encoding the time-bin on it, and performing a projective measurement on the meter that should reveal the time-bin. Alice and Bob are free to choose the initial ancilla state. In this section we assume that Alice and Bob cannot share entanglement, therefore the meter state must be separable. We will take the initial meter state to be a pure, separable state

\begin{equation} \label{eq:meter_initial_state}
    \chi_0 = \ket{\bar{0}}\bra{\bar{0}}, \quad \ket{\bar{0}} = \ket{0}_{A,m} \otimes \ket{0}_{B,m},
\end{equation}
where the $A$ and $B$ indices denote the states received by Alice and Bob, and $m$ denotes the meter. The initial state of the total system is

\begin{equation}
    \rho = \rho_r \otimes \chi_0.
\end{equation}

To measure the time-bin, Alice and Bob interact the referee state with the meter by a unitary operator $U$. They wish that the interaction has the following form

\begin{equation} \label{eq:desired_state}
\begin{aligned}
    \rho' & = U \rho U^\dagger \\
    & = \int_{\phi_0}^{\phi_0+2\pi}d\phi\sum_{n=0}^N p(n,\phi) U \left[ \ket{\Psi_{\phi,n}} \bra{\Psi_{\phi,n}} \otimes \chi_0 \right] U^\dagger \\
    & = \int_{\phi_0}^{\phi_0+2\pi}d\phi\sum_{n=0}^N p(n,\phi) \ket{\Psi_{\phi,n}} \bra{\Psi_{\phi,n}} \otimes \chi_n,
\end{aligned}
\end{equation}
where $\{ \chi_n \}$ are orthogonal meter states. One can understand (\ref{eq:desired_state}) as the weighted sum of of different possible pure states. Note that in the second line of (\ref{eq:desired_state}) the term in the square brackets is a pure state $\ket{\Psi_{\phi,n}}\otimes \ket{\bar{0}}$. Therefore, $U$ acting on that term should return a pure state as well. We expect that $U$ should perform the following operation

\begin{equation} \label{eq:desired_state_2}
    U \ket{\Psi_{\phi,n}}\otimes \ket{\bar{0}} = \ket{\Psi_{\phi,n}}\otimes \ket{\bar{n}}
\end{equation}
with $\braket{\bar{n} | \bar{m}} = \delta_{\bar{n} \bar{m}}$, $\{ \ket{\bar{n}} \}$ are the orthogonal meter states. We observe that the state on the right is separable.
After such interaction, Alice and Bob would measure the meter in the $\{ \ket{\bar{n}} \}$ basis to recover the time-bin. 

$U$ represents unitary operations performed by Alice and Bob locally within their laboratories; hence, it must have the form
\begin{equation} \label{eq:U}
    U = U_A \otimes U_B
\end{equation}
where $U_{A}$ $(U_{B})$ acts only on Alice's (Bob's) states. Consider the $n=0$ case in (\ref{eq:desired_state_2}) corresponding to the referee sending the vacuum within all the time-bins. It is useful to express the corresponding referee state $\ket{\Psi_{\phi,0}}$ using the same notation as in: (\ref{eq:ref_bell_pair})
\begin{align} \label{eq:n0}
     \ket{\Psi_{\phi,0}} = \ket{0,k}_{A,r} \ket{0,k}_{B,r}
    \end{align}
with $k$ being an arbitrary integer $k\in\{ 1,...,N \} $.
To keep track which quantum states come from the referee, we introduce the index $r$.
We use (\ref{eq:U}) and (\ref{eq:meter_initial_state})
\begin{equation}
\begin{aligned}
    U & \ket{\Psi_{\phi,0}}\otimes \ket{\bar{0}} \\
    & = U_A \ket{0,k}_{A,r}\ket{0}_{A,m} \otimes \ U_B \ket{0,k}_{B,r}\ket{0}_{B,m} \\
    & = \ket{0,k}_{A,r}\ket{0}_{A,m} \otimes \  \ket{0,k}_{B,r}\ket{0}_{B,m} \\
    & = \ket{\Psi_{\phi,0}} \otimes \ket{\bar{0}}
\end{aligned}
\end{equation}
where the final two lines are the right-hand side of Eq. (\ref{eq:desired_state_2}). The equation above implies
\begin{equation} \label{eq:n0_2}
\begin{aligned}
    U_A \ket{0,k}_{A,r} \ket{0}_{A,m} & = \ket{0,k}_{A,r} \ket{0}_{A,m}, \\
    U_B \ket{0,k}_{B,r} \ket{0}_{B,m} & = \ket{0,k}_{B,r} \ket{0}_{B,m}.
\end{aligned}
\end{equation}

Now consider the $n>0$ case 

\begin{equation}
\begin{aligned}
    & U  \ket{\Psi_{\phi,n}}\otimes \ket{\bar{0}} \\
    & = \frac{1}{\sqrt{2}} U_A \ket{1,n}_{A,r}\ket{0}_{A,m} \otimes \ U_B \ket{0,n}_{B,r}\ket{0}_{B,m} \\
    & + \frac{e^{i\phi}}{\sqrt{2}} U_A \ket{0,n}_{A,r}\ket{0}_{A,m} \otimes \ U_B \ket{1,n}_{B,r}\ket{0}_{B,m} \\
    & = \frac{1}{\sqrt{2}} U_A \ket{1,n}_{A,r}\ket{0}_{A,m} \otimes \ \ket{0,n}_{B,r}\ket{0}_{B,m} \\
    & + \frac{e^{i\phi}}{\sqrt{2}} \ket{0,n}_{A,r}\ket{0}_{A,m} \otimes \ U_B \ket{1,n}_{B,r}\ket{0}_{B,m}
\end{aligned}
\end{equation}
where in the final line we used the results from (\ref{eq:n0_2}). We observe that unless $U_A$ and $U_B$ leave the meter state unchanged, the final meter state will not remain separable, as it was the case in (\ref{eq:desired_state_2}). However, if $U_A$ and $U_B$ leave the meter state unchanged, then the ancilla would lose its purpose as a meter, since it would not have the time-bin encoded in it. We conclude that achieving the desired meter operation (\ref{eq:desired_state}), and therefore winning the clock game, is impossible if we do not allow for the shared entanglement in the ancilla. The amount of shared entanglement needed to win the game was discussed in Sec. \ref{sec:LOSE}. In the next appendix section we demonstrate that we can consider the entanglement cost based on the properties that the ancilla must satisfy to serve as a reliable measurement device.

\section{Necessity of the Communication with the Referee in the Clock Game} \label{sec:classical_communication}

The final step of the clock game contains the communication of classical information: two integers, $x$ and $y$, based on which the referee can recover the time-bin within which she has sent the excitation. One can ask whether it is possible for both parties to recover the time-bin locally, i.e., whether both parties can extract the information about the time bin based on locally available resources, which are the ancilla and the referee's state. 

We will assume that the meter has to obey the rule (\ref{eq:desired_state}) which implies that (\ref{eq:desired_state_2}) must be satisfied. However, now we will not assume that $\ket{\bar{0}}$ must be a separable state. Let the basis states for Alice's and Bob's ancilla be $\{\ket{p}_{A,m} \}$ and for $\{\ket{q}_{B,m} \}$. We take the initial state of the meter to be 

\begin{equation}
    \ket{\bar{0}} = \sum_{p,q} c_{p,q}^{(0)} \ket{p}_{A,m} \ket{q}_{B,m}.
\end{equation}

After the interaction of the referee's state and the meter, the meter should have the time-bin encoded in it. We take the meter's state with encoded time-bin to be 
\begin{equation}
    \ket{\bar{n}} = \sum_{p,q} c_{p,q}^{(n)} \ket{p}_{A,m} \ket{q}_{B,m}.
\end{equation}
We will keep the general form of the coefficients $c_{p,q}^{(n)}$, which should be chosen to satisfy $\braket{\bar{n}|\bar{m}} = \delta_{n,m}$. Evaluating (\ref{eq:desired_state_2}) for $n=0$ results in 

\begin{equation}
\begin{aligned}
    & U  \ket{\Psi_{\phi,0}}\otimes \ket{\bar{0}} \\
    & = U_A \otimes U_B \ket{0,k}_{A,r} \ket{0,k}_{B_r} \sum_{p,q} c_{p,q}^{(0)} \ket{p}_{A,m} \ket{q}_{B,m} \\
    & = \sum_{p,q} c_{p,q}^{(0)} \ U_A \ket{0,k}_{A,r} \ket{p}_{A,m} \otimes U_B \ket{0,k}_{B,r} \ket{q}_{B,m} \\
    & = \sum_{p,q} c_{p,q}^{(0)} \ket{0,k}_{A,r} \ket{p}_{A,m} \otimes \ket{0,k}_{B,r} \ket{q}_{B,m}
\end{aligned}
\end{equation}
where the final line is the right-hand side of (\ref{eq:desired_state_2}). The equation above implies the following rule on the local unitaries $U_A$ and $U_B$:
\begin{equation} \label{eq:U_AB_rule}
\begin{aligned}
    U_A \ket{0,k}_{A,r} \ket{p}_{A,m} & = \ket{0,k}_{A,r} \ket{p}_{A,m} \\
    U_B \ket{0,k}_{B,r} \ket{q}_{B,m} & = \ket{0,k}_{B,r} \ket{q}_{B,m},
\end{aligned}
\end{equation}
i.e., the local meter's state remains unchanged if no excitation arrived from the referee. Consider now (\ref{eq:desired_state_2}) for $n>0$:
\begin{equation} \label{eq:step1}
\begin{aligned}
    & U \ket{\Psi_{\phi,n}} \ket{\bar{0}} \\
    & = \frac{1}{\sqrt{2}} \sum_{p,q} c_{pq}^{(0)} U_A \ket{1,n}_{A,r} \ket{p}_{A,m} \  U_B \ket{0,n}_{B,r} \ket{q}_{B,m} \\
    & + \frac{e^{i\phi}}{\sqrt{2}} \sum_{p,q} c_{pq}^{(0)} U_A \ket{0,n}_{A,r} \ket{p}_{A,m} \  U_B \ket{1,n}_{B,r} \ket{q}_{B,m} \\
    & = \frac{1}{\sqrt{2}}\sum_{p,q} c_{pq}^{(0)} U_A \ket{1,n}_{A,r} \ket{p}_{A,m} \  \ket{0,n}_{B,r} \ket{q}_{B,m} \\
    & + \frac{e^{i\phi}}{\sqrt{2}} \sum_{p,q} c_{pq}^{(0)} \ket{0,n}_{A,r} \ket{p}_{A,m} \  U_B \ket{1,n}_{B,r} \ket{q}_{B,m}
\end{aligned}
\end{equation}
where we used (\ref{eq:U_AB_rule}). According to Eq. (\ref{eq:desired_state_2}), the referee's state remains unchanged after the interaction with the meter. Therefore, we assume that the local unitaries follow
\begin{align}
    U_A \ket{1,n}_{A,r} \ket{p}_{A,m} & = \ket{1,n}_{A,r} U_A^{(n)}, \ket{p}_{A,m} \\
    U_B \ket{1,n}_{B,r} \ket{q}_{B,m} & = \ket{1,n}_{B,r} U_B^{(n)} \ket{1}_{B,m}.
\end{align}
The unitaries $U_A^{(n)}$ and $U_B^{(n)}$ act only on the local meter state. We indicate by superscript $(n)$ that they are time-bin dependent. The equation above states that if the excitation arrived within time-bin $n$, it is encoded on the meter by the unitary $U_{A}^{(n)}$ $(U_{B}^{(n)})$. Apply these rules to (\ref{eq:step1})

\begin{equation}
\begin{aligned}
    & U \ket{\Psi_{\phi,n}} \ket{\bar{0}} \\
    & = \frac{1}{\sqrt{2}}\sum_{p,q} c_{pq}^{(0)} \ket{1,n}_{A,r} U_A^{(n)} \ket{p}_{A,m} \  \ket{0,n}_{B,r} \ket{q}_{B,m} \\
    & + \frac{e^{i\phi}}{\sqrt{2}} \sum_{p,q} c_{pq}^{(0)} \ket{0,n}_{A,r} \ket{p}_{A,m} \ \ket{1,n}_{B,r} U_B^{(n)} \ket{q}_{B,m} \\
    & = \frac{1}{\sqrt{2}}\ket{1,n}_{A,r} \ket{0,n}_{B,r} \sum_{p,q} c_{pq}^{(0)} U_A^{(n)} \ket{p}_{A,m}\ket{q}_{B,m} \\
    & + \frac{e^{i\phi}}{\sqrt{2}} \ket{0,n}_{A,r} \ket{1,n}_{B,r} \sum_{p,q} c_{pq}^{(0)} \ket{p}_{A,m} U_B^{(n)} \ket{q}_{B,m}. 
\end{aligned}
\end{equation}
According to (\ref{eq:desired_state_2}), (\ref{eq:step1}) should return a separable state of the referee's state and the ancilla. It requires
\begin{equation} \label{eq:step2}
\begin{aligned}
    &\sum_{p,q} c_{pq}^{(0)} U_A^{(n)} \ket{p}_{A,m}\ket{q}_{B,m} \\
    = & \sum_{p,q} c_{pq}^{(0)} \ket{p}_{A,m} U_B^{(n)} \ket{q}_{B,m} \\
    = & \sum_{p,q} c_{pq}^{(n)} \ket{p}_{A,m} \ket{q}_{B,m} = \ket{\bar{n}},
\end{aligned}
\end{equation}
where the final line follows from the right hand side of (\ref{eq:desired_state_2}). Rewrite the first line of (\ref{eq:step2})
\begin{equation} \label{eq:line1}
\begin{aligned}
    & \sum_{p,q} c_{pq}^{(0)} U_A^{(n)} \ket{p}_{A,m}\ket{q}_{B,m} \\
    = & \left[ U_A^{(n)} \otimes \mathbb{1}_B \right] \sum_{p,q} c_{pq}^{(0)} \ket{p}_{A,m}\ket{q}_{B,m} \\
    = & \left[ U_A^{(n)} \otimes \mathbb{1}_B \right] \ket{\bar{0}} = \ket{\bar{n}}.
\end{aligned}
\end{equation}
Similarly, for the second line of (\ref{eq:desired_state_2}) we get
\begin{equation} \label{eq:line2}
\begin{aligned}
    & \sum_{p,q} c_{pq}^{(0)} \ket{p}_{A,m} U_B^{(n)} \ket{q}_{B,m} \\
    = & \left[ \mathbb{1}_A \otimes U_B^{(n)} \right] \sum_{p,q} c_{pq}^{(0)} \ket{p}_{A,m}\ket{q}_{B,m} \\
    = & \left[ \mathbb{1}_A \otimes U_B^{(n)} \right] \ket{\bar{0}} = \ket{\bar{n}}.
\end{aligned}    
\end{equation}
Equations (\ref{eq:line1}) and (\ref{eq:line2}) imply that one must be able to transform the ancilla state from $\ket{\bar{0}}$ to $\ket{\bar{n}}$ just by performing a local operation either in Alice's or Bob's laboratory. 

We are now ready to consider whether or not it is possible to win the clock game without the communication with the referee. To achieve it, Alice and Bob must be able to determine the time-bin just based on locally available resources, without the classical communication between each other or with the referee. They are also not allowed to establish a quantum channel between them other that the ancilla state.
They perform the measurements on the ancilla (meter) quantum state, which must have the time-bin encoded in it. However, a stronger condition is needed if one wants to determine the time-bin locally: both local ancilla quantum states must have the time-bin encoded in it. 

Let us assume that the locally available ancilla quantum states have $D_A$ (Alice's meter) and $D_B$ (Bob's meter) orthogonal levels. Local time-bin measurement requires that the local measurements performed on these states must return the information about the time-bin. Therefore, we divide these levels into groups and assign the corresponding time-bins to them. For Alice, we denote the basis in which she performs the measurement on the meter in the following way
\begin{equation}
\begin{aligned}
    \{ & \ket{0_1}_{A,m},\ket{0_2}_{A,m},...,\ket{0_{n_{0}}}_{A,m}, \\
    & \ket{1_1}_{A,m},\ket{1_2}_{A,m},...,\ket{1_{n_{1}}}_{A,m},    \\
    & ... \\
    & \ket{N_1}_{A,m},\ket{N_2}_{A,m},...,\ket{N_{n_{N}}}_{A,m} \}
\end{aligned}
\end{equation}
where $n_i$ is the number of levels assigned to time-bin $i$. After the local processing Alice should be able to measure her local ancilla state in this basis to obtain the time-bin: the result $\ket{j_k}_{A,m}$ corresponds to time-bin $j$. One can define a similar basis for Bob. 

Before the measurement the meter must have $n=0$ encoded in it not only globally, but also locally. The most general form of the meter state that satisfies it is
\begin{equation}
    \ket{\bar{0}} = \sum_{i,j} c_{ij} \ket{0_i} \ket{0_j}.
\end{equation}
It is an entangled state, as required by the results of Appendix \ref{sec:entanglement}. From (\ref{eq:line1}) and (\ref{eq:line2}) we know, that transforming that state to $\ket{\bar{n}}$ must be possible only by performing local operations in only one of the laboratories. If Alice is the one to perform such operation, then
\begin{equation}
\begin{aligned}
    \left[ U_A^{(n)} \otimes \mathbb{1} \right] \ket{\bar{0}} & = \sum_{i,j} c_{ij} \ U_A^{(n)} \ket{0_i} \ket{0_j} \\
    & = \sum_{i,j} c_{ij}  \ket{n_i} \ket{0_j}
\end{aligned}
\end{equation}
which encodes the time-bin on only one of the local states. A similar argument can be applied for Bob. We conclude that the rules (\ref{eq:line1}) and (\ref{eq:line2}) prevent one from encoding the time-bin on both locally available ancilla states. Therefore, the desired meter operation cannot be achieved and one cannot win the clock game based only on the locally available resources. Classical communication between the parties, or with the referee, is required. 

\section{Entanglement and dimensionality of the ancilla as a resource} \label{sec:ent_dim}

In Appendix \ref{sec:entanglement} we have shown that one cannot win the clock game without an entangled resource, but we have not determined the degree of entanglement needed to succeed. We consider it in this appendix together with the dimensionality of the local ancilla systems needed to win the game. First, we examine the simplest nontrivial case of $N=1$ where the referee can send the phase encoded state in one time-bin. It will help us to establish important concepts needed for more general case of arbitrary number of time-bins. 

Let us consider the needed dimensionality of the ancilla systems needed to win the clock game for the $N=1$ case. Naturally, allowing the local ancilla states to have only one level is not enough, since then it is impossible to use it as a meter that verifies the presence of the phase encoded state. Therefore, the smallest nontrivial number of levels to consider is two. In this appendix, we work in the meter basis in which Alice and Bob perform the measurement, with the possible measurement results being
$\{ \ket{0}_A \ket{0}_B,   \ket{0}_A \ket{1}_B,  \ket{1}_A \ket{0}_B,  \ket{1}_A \ket{1}_B \}$. We omit the index $m$ denoting the meter, since in this appendix we work only with the meter quantum states. 

The initial meter state must be an entangled state that must have time-bin $0$ encoded in it. Therefore, it must be constructed from at least two of the kets from the measurement basis. We are free to choose these kets, but we must remember that they cannot allow for local encoding of the time-bin. Therefore, the choice of $\ket{0}_A \ket{0}_B$ and $\ket{0}_A \ket{1}_B$ is not allowed since it encodes time-bin $0$ on Alice's state. 

Define the space 
\begin{equation} \label{eq:S0}
    \mathcal{S}_0 = \{ \ket{0}_A \ket{0}_B , \ket{1}_A \ket{1}_B \}
\end{equation}
which contains the vectors assigned to time-bin $0$. The remaining vectors are assigned to time-bin $1$ space
\begin{equation} \label{eq:S1}
    \mathcal{S}_1 = \{  \ket{0}_A \ket{1}_B,  \ket{1}_A \ket{0}_B \}.
\end{equation}
Note that the assignment of vectors to the time-bin spaces is based on the parity of the vectors. Other assignment is also allowed ($\mathcal{S}_0 \leftrightarrow \mathcal{S}_1$), which is equivalent to relabeling the local Alice's, or Bob's, states according to $\ket{0}_{A(B)} \leftrightarrow \ket{1}_{A(B)} $. We will continue with the choice (\ref{eq:S0}) and (\ref{eq:S1}). Note that if the measurement is performed, both local measurement results are required to establish the time-bin.

The general pure state with time-bin $0$ encoded in it is
\begin{equation} \label{eq:initial_state}
    \ket{\phi_0} = c_0 \ket{0}_A \ket{0}_B + c_1 \ket{1}_A \ket{1}_B.
\end{equation}
According to (\ref{eq:line1}) and (\ref{eq:line2}), encoding time-bin in it should be possible only by performing local operations on one of the local quantum states. The general form of local operations that achieves it is 
\begin{equation} \label{eq:local_operations}
\begin{aligned}
    & U_A \ket{0}_A = e^{i\alpha_0}\ket{1}_A, &  U_A \ket{1}_A = e^{i\alpha_1}\ket{0}_A, \\
    & U_B \ket{0}_B = e^{i\beta_0}\ket{1}_B, &  U_B \ket{1}_B = e^{i\beta_1}\ket{0}_B,    
\end{aligned}
\end{equation}
i.e., the local operations behave like the X gates in the measurement basis up to a phase factor. We have omitted the superscript $n$ in the local unitaries, since for one allowed time-bin there is only one value of $n$ for which the local unitaries are not identity operations. Let us encode time-bin $1$ on the state (\ref{eq:initial_state}) by applying the local operation $U_A$ on Alice's state
\begin{equation} \label{eq:final1}
\begin{aligned}
   \left[ U_A \otimes \mathbb{1}_B \right] \ket{\phi_0}
   & = e^{i\alpha_0} c_0 \ket{1}_A \ket{0}_B \\
   & + e^{i\alpha_1} c_1 \ket{0}_A \ket{1}_B. 
\end{aligned}
\end{equation}
Now encode the same time-bin by applying the local operation $U_B$ on Bob's state
\begin{equation} \label{eq:final2}
\begin{aligned}
    \left[ \mathbb{1}_A \otimes U_B\right] \ket{\phi_0} 
    & = e^{i\beta_0} c_0 \ket{0}_A \ket{1}_B \\
    & + e^{i\beta_1} c_1 \ket{1}_A \ket{0}_B.    
\end{aligned}
\end{equation}
Both states (\ref{eq:final1}) and (\ref{eq:final2}) belong to the space $\mathcal{S}_1$, as they should. The results (\ref{eq:line1}) and (\ref{eq:line2}) imply that they must be equal to each other, which requires 
\begin{equation}
    e^{i\alpha_0} c_0 = e^{i\beta_1} c_1, \ e^{i\beta_0} c_0 = e^{i\alpha_1} c_1.
\end{equation}
Taking the absolute value of both sides of any of these equations results in 
\begin{equation}
    |c_0| = |c_1|,
\end{equation}
implying that (\ref{eq:initial_state}) is a maximally entangled state. It establishes that to examine 1 time-bin one needs $2$-dimensional local ancilla states in a maximally entangled state, which is the case in the clock game winning strategy discussed in Sec. \ref{sec:pairwise_1pair}. 

Let us now examine the case of $N\geq 1$ allowed time-bins within which the referee can provide the phase encoded state. As before, we will work in the meter measurement basis. Let us pick one of the states from that basis, $\ket{0}_A \ket{0}_B$, and assign it to the time-bin $0$ space $\mathcal{S}_0$. Note that the choice of state does not change our argument, e.g., if one picks the $\ket{2}_A \ket{7}_B$, then one can just relabel the local states $\ket{2}_A \rightarrow \ket{0}_A$, $\ket{7}_B \rightarrow \ket{0}_B$. 

For $N$ time-bins we define $N+1$ spaces $\mathcal{S}_0,\mathcal{S}_1,...,\mathcal{S}_N$; each of them assigned to corresponding time-bin. The local operations performed on only one of the meter states should allow one to take any state from $\mathcal{S}_0$, and take it to other desired space. For example, $U_A^{(n)}$ operation applied on Alice's state should take the global meter state from $\mathcal{S}_0$ to the $\mathcal{S}_n$ space. In particular, it should apply to the $\ket{0}_A \ket{0}_B \in \mathcal{S}_0$ state. 

Let us assign the states $\ket{n}_A \ket{0}_B$ and $\ket{0}_A \ket{n}_B$ to time-bin $n$ with corresponding space $\mathcal{S}_n$. Note that it does not result in the loss of generality since one can compensate for other assignment of the vector space by relabeling the local states. For example, if one assigns the vector $\ket{3}_A \ket{0}_B$ to the $\mathcal{S}_5$ space, then we can relabel $\ket{3}_A \rightarrow \ket{5}_A $ to come back to the initial choice (other states might have to be relabeled to compensate for that change).

Then, similarly to (\ref{eq:local_operations}), the local unitaries should affect the local states according to
\begin{equation} \label{eq:local_unitaries2}
\begin{aligned}
    U_A^{(n)}\ket{p}_A & = e^{i\alpha_p}\ket{p+1 \text{ mod } N+1}_A \\
    U_B^{(n)}\ket{p}_B & = e^{i\beta_p}\ket{p+1 \text{ mod } N+1}_B,
\end{aligned}
\end{equation}
which results in 
\begin{equation}
\begin{aligned}
    \left[ U_A^{(n)} \otimes \mathbb{1}_B \right] \ket{0}_A \ket{0}_B 
    & = e^{i\alpha_p} \ket{n}_A \ket{0}_B \in \mathcal{S}_n \\
    \left[ \mathbb{1}_A \otimes U_B^{(n)} \right] \ket{0}_A \ket{0}_B 
    & = e^{i\beta_p} \ket{0}_A \ket{n}_B \in \mathcal{S}_n.
\end{aligned}
\end{equation}
Since the unitaries $U_{A(B)}^{(n)}$ result in a set of distinguishable results for different $n$'s, the sets $\{\ket{n}_A \ket{0}_B, n = 0,...,N\} $ and $\{\ket{0}_A \ket{n}_B, n = 0,...,N\} $ must both contain $N+1$ orthogonal vectors. It is achieved only if the local meter systems have at least $N+1$ distinguishable levels.

Given that we know the dimension of the local meter states, we are ready to assign them to the time-bin spaces. It must be done in such a way that transforming a state assigned to time-bin $0$ (space $\mathcal{S}_0$) to time-bin $n$ (space $\mathcal{S}_n$) is possible only by performing local operations with the restriction that the time-bin cannot be assigned locally. It is achieved by the following assignment:

\begin{equation}
\begin{aligned}
     \mathcal{S}_n = 
     & \{ \text{all states } \ket{p}_A \ket{q}_B \\
     & \text{ for which }p+q \text{ mod }N+1 = n \}   
\end{aligned}
\end{equation}
Other allowed assignments are equivalent, since they are achieved by relabeling the local states. The general form of a pure state belonging to $\mathcal{S}_0$ is
\begin{equation} \label{eq:initialN}
\begin{aligned}
    \ket{\phi_0^{(N)}} = \sum_{p=0}^N
    & c_p \ket{p \text{ mod } N+1}_A \\
    & \otimes \ket{-p \text{ mod } N+1}_B \in \mathcal{S}_0   
\end{aligned}
\end{equation}

Let us modify the state (\ref{eq:initialN}) and encode time-bin $n$ in it by applying the local unitary (\ref{eq:local_unitaries2}) on Alice's state

\begin{equation} \label{eq:output1}
\begin{aligned}
    & \left[ U_A^{(n)} \otimes \mathbb{1}_B \right] \ket{\phi_0^{(N)}}= \sum_{p=0}^N c_p e^{i\alpha_p} \times \\
    & \times \ket{p+n \text{ mod } N+1}_A 
    \ket{-p \text{ mod } N+1}_B  \in \mathcal{S}_n
\end{aligned}
\end{equation}  
Now encode the same time-bin by applying the unitary on Bob's state
\begin{equation} \label{eq:output2}
\begin{aligned}
    & \left[ \mathbb{1}_A \otimes U_B^{(n)} \right] \ket{\phi_0^{(N)}}= \sum_{p=0}^N c_p e^{i\beta_p} \times \\
    & \times \ket{p \text{ mod } N+1}_A 
    \ket{-p+n \text{ mod } N+1}_B  \in \mathcal{S}_n
\end{aligned}
\end{equation}
According to (\ref{eq:final1}) and (\ref{eq:final2}), equations (\ref{eq:output1}) and (\ref{eq:output2}) should result in the same state. It is achieved by making all the coefficients in front of the same kets equal to each other. For all allowed values of $p$ one gets
\begin{equation}
    c_p e^{i\alpha_p} = c_{p'}e^{i\alpha_{p'}}, \ p' = p+1 \text{ mod }N+1.
\end{equation}
It implies
\begin{equation}
    |c_p| = |c_{p'}| \text{ for }p=0,1,...,N,
\end{equation}
i.e., the absolute values of all the $c_p$ coefficients in the state (\ref{eq:initialN}) must be equal to each other. Therefore, the initial state of the meter must be a maximally entangled state.

\end{document}